\documentclass[a4paper,11pt]{article}
\pdfoutput=1 

\usepackage{jheppub} 
\usepackage{amsmath}  
\usepackage{physics}  
\usepackage{slashed}  
\usepackage{braket}
\usepackage{amssymb}
\usepackage{bbm}
\usepackage{bm}
\usepackage[makeroom]{cancel}
\usepackage{empheq}
\usepackage{mathtools}
\usepackage{cancel}
\usepackage{comment}
\usepackage{xcolor}
\usepackage{multirow}
\usepackage{enumitem}
\usepackage{tabularx}
\usepackage[normalem]{ulem}
\usepackage{cleveref}
\usepackage{xspace}

\makeatletter
\gdef\@fpheader{}
\g@addto@macro\bfseries{\boldmath}
\makeatother

\newcommand{\beq}{\begin{equation}}
\newcommand{\eeq}{\end{equation}}
\newcommand{\bea}{\begin{equation}\begin{aligned}}
\newcommand{\eea}{\end{aligned}\end{equation}}

\newcommand{\sss}[1]{{\scriptscriptstyle{#1}}}

\newcommand{\Mp}{M_{\sss{\mathrm{Pl}}}}
\newcommand{\ie}{i.e.\xspace} 
\newcommand{\eg}{e.g.\xspace} 
\newcommand{\cs}{c_{\sss{\mathrm{S}}}}
\newcommand{\alphaR}{\alpha_{\mathrm{R}}}
\newcommand{\alphaI}{\alpha_{\mathrm{I}}}
\newcommand{\betaR}{\beta_{\mathrm{R}}}
\newcommand{\betaI}{\beta_{\mathrm{I}}}

\newcommand{\epsOnestar}{\epsilon_{1\star}}

\title{When inflationary perturbations refuse to classicalise: the role of non-Gaussianity in Wigner negativity}

\author[a]{Aurora Ireland,}
\author[b]{Vincent Vennin}

\affiliation[a]{Leinweber Institute for Theoretical Physics, Stanford University, Stanford, CA 94305, USA}
\affiliation[b]{Laboratoire de Physique de l'Ecole Normale Sup{\'e}rieure, ENS, CNRS, Universit{\'e} PSL, Sorbonne Universit{\'e}, Universit{\'e} Paris Cit{\'e}, 75005 Paris, France}

\emailAdd{anireland@stanford.edu}
\emailAdd{vincent.vennin@phys.ens.fr}

\abstract{Inflationary perturbations are quantum in origin. Yet, when computing cosmological observables, they are often treated as classical stochastic fields. Do they nevertheless retain quantum birthmarks?
A hallmark of genuinely quantum behaviour is quantum interferences, arising from phase coherence between distinct branches of the wavefunction. Such interference is diagnosed by the non-positivity of the Wigner function, and according to Hudson’s theorem, the only pure states with positive Wigner functions are Gaussian states. Consequently, any departure from Gaussianity necessarily implies a non-positive Wigner function, precluding a description in terms of a classical distribution.
This motivates us to compute the Wigner function of curvature perturbations, accounting for primordial non-Gaussianities, using the EFT of inflation. We find that the Wigner function develops pronounced interference fringes on super-Hubble scales, and in particular its negativity grows as $a^2$ in ultra-slow-roll backgrounds.
These results demonstrate that quantum effects can remain significant at late times, and that squeezing alone does not ensure classicality, contrary to standard lore. This suggests that the prospects for detecting genuinely quantum signatures of the universe's origins in cosmological observables may be less bleak than previously thought.
}

\begin{document}
\maketitle
\flushbottom

\section{Introduction}\label{sec:intro}

The leading paradigm for the formation of cosmic structure in our universe posits that it originates from quantum fluctuations of the vacuum during an early epoch of inflation~\cite{Starobinsky:1980te,Sato:1981,Guth:1980zm,Linde:1981mu,Albrecht:1982wi}. As the comoving Hubble horizon shrinks, these fluctuations are stretched to super-Hubble scales, where they undergo parametric amplification. Later, following the end of inflation and a period of reheating, the perturbations re-enter the horizon during either radiation or matter domination, subsequently evolving into the density perturbations that seed the large-scale structure observed today. The predicted spectrum and statistical properties of these perturbations~\cite{Starobinsky:1979ty,Mukhanov:1981xt,Guth:1982ec,Starobinsky:1982ee,Bardeen:1983} are in exceptional agreement with observations of the cosmic microwave background (CMB) anisotropies~\cite{Planck:2018jri}. 

In this picture, inflationary perturbations are fundamentally quantum in origin. In practice, however, there are many contexts in which it is appropriate to treat these quantum fields as classical stochastic variables. In particular, two-point correlation functions can always be reproduced with a stochastic description~\cite{Martin:2015qta}: one simply needs to consider a Gaussian random field whose covariance matrix matches the quantum two-point expectation values of the underlying quantum field. A standard approach in cosmology is thus to start from stochastic, Gaussian initial conditions at the end of inflation and evolve them as classical fields afterwards, according to the equations of cosmological perturbation theory. This standard practice raises two issues, however. First, at the practical level, it is not guaranteed that it properly captures  higher-point statistics. Second, at the conceptual level, how a definite classical field configuration emerges from a quantum superposition of states remains elusive (this is the so-called quantum measurement problem).

\subsection*{Quantum-to-classical transition}

The usual way to sidestep both problems and justify the replacement of quantum operators by classical stochastic fields is to assume that some sort of \textit{quantum-to-classical transition} occurs at super-Hubble scales during inflation.\footnote{We emphasise that this transition is not intended as a resolution to the quantum-measurement problem in general, but rather as an effective criterion used in cosmology to motivate the replacement of quantum operators by stochastic fields. In particular, decoherence does not solve the quantum measurement problem~\cite{Sudarsky:2009, Martin:2012}.} Note that this transition is also essential to the stochastic approach to inflation~\cite{Starobinsky:1986fx}, which provides an effective description of the long-wavelength modes by modelling the influence of shorter-wavelength fluctuations as an effective noise term. There are two main arguments usually invoked in support of such a transition.

First, on super-Hubble scales the Heisenberg-picture curvature perturbation operator can be decomposed into a frozen mode and a decaying mode. The commutator between the curvature perturbation and its conjugate momentum, when compared with the anti-commutator, scales as the decaying mode, and hence becomes strongly suppressed. This has been argued to render the field and its momentum approximately commuting, and thus ``classical''~\cite{Guth:1985ya, Polarski:1995,Lesgourgues:1996,Kiefer:1998qe,Kiefer:2008,dePutter:2019xxv}.\footnote{This is sometimes referred to as ``decoherence without decoherence'', although we shall avoid such terminology since cosmological perturbations are described by a pure (\ie~non-decohered) state in this approach.} In the Schr\"odinger picture, this corresponds to quantum squeezing~\cite{Grishchuk:1990bj}: as the quantum state becomes localised along one phase-space direction, the squeezed direction becomes too suppressed to be measured, and the commutator cannot be accessed. However, this criterion is not invariant under field reparametrisations: by performing a phase-space rotation, the anti-commutator between the field and its momentum can always be made zero, contrary to the commutator whose size is fixed by $\hbar$. More generally, squeezing can always be undone by an appropriate linear canonical transformation~\cite{Grain:2019vnq}, and hence it does not capture intrinsic (in technical terms, symplectic-invariant) properties of the quantum state~\cite{Agullo:2022ttg}.

Second, curvature perturbations do not form an isolated system: they may interact with other fields present in the early universe, becoming entangled with them. Even in the absence of other fields, gravitational self-interactions inevitably couple observable scales to inaccessible, small-wavelength modes~\cite{Burgess:2022nwu}. There also exists entanglement in real space between the spatial regions inside and outside our observable horizon~\cite{Sharman:2007gi}. When tracing over these environmental degrees of freedom, curvature perturbations need to be described by an open quantum system, subject to decoherence~\cite{Joos:1984}. In the ``pointer basis'' selected by the interaction, quantum interferences become strongly suppressed~\cite{Zurek:1981}, and the system effectively behaves as a statistical mixture of states, rather than a coherent superposition~\cite{Brandenberger:1990bx,Lombardo:2005iz,Burgess:2006jn,Martineau:2006ki,Sharman:2007gi,Nelson:2016kjm,Shandera:2017qkg,Martin:2018zbe,Martin:2018lin,Martin:2021znx,Colas:2022hlq,Colas:2022kfu,DaddiHammou:2022itk,Burgess:2022nwu,Lopez:2025arw}. Even if decoherence is effective, it does not necessarily imply that all quantum features are erased: as shown in~\cite{Martin:2021znx}, what matters is the \emph{rate} at which decoherence proceeds, rather than the final amount.\footnote{Concretely, in inflating spacetimes for decoherence to erase self-entanglement within curvature perturbations, purity needs to decay faster than $a^{-4}$, where $a$ is the scale factor.} Moreover, commonly used diagnostics (such as the purity) depend on the choice of coarse graining underlying the system-environment split.

\subsection*{Can we prove that cosmic structures have a quantum-mechanical origin?}


Setting aside the question of whether the above transition provides a satisfactory account of classicality, one may instead ask a more operational question: can signatures of an underlying quantum origin persist in late-time observables? There have been several attempts to reveal the presence of genuine quantum features in primordial perturbations. These range from calculating entanglement measures such as quantum discord~\cite{Lim:2014uea, Martin:2015qta}, to constructing Bell~\cite{Campo:2005sv, Maldacena:2015bha, Martin:2016tbd, Choudhury:2016cso, Martin:2017zxs, Kanno:2017dci} or Leggett-Garg~\cite{Martin:2016nrr,Ando:2020kdz} inequalities, to searching for features in the primordial bispectrum~\cite{Green:2020whw, Green:2022fwg,Launay:2024trh,Belfiglio:2025exd}.

Ultimately, this is a notoriously difficult task for at least two logically independent reasons. Dynamically, decoherence suppresses phase-sensitive information in the reduced state, making quantum signatures harder to access in late-time observables.
A second difficulty is that the quantum correlations generated during inflation are primarily encoded between Fourier modes, which does not lead to significant entanglement between disjoint regions of physical space~\cite{Martin:2021xml, Martin:2021qkg, Espinosa-Portales:2022yok, Agullo:2023fnp, Boutivas:2023mfg, Belfiglio:2025cst}. This is because local measurements of quantum fields implicitly involve a trace over unobserved regions, which leads to an effective decoherence in the spatially local description that suppresses accessible entanglement.\footnote{This can be further interpreted as delocalisation of the partner mode~\cite{Agullo:2024cln,RibesMetidieri:2025ets,Ribes-Metidieri:2025nfw,Agullo:2025dxp}.}


\subsection*{Quantum interference and the Wigner function}

For this reason, rather of attempting to reveal entanglement, in this work we focus on another hallmark of quantum theory: the presence of interferences. In phase space, quantum interferences are evidenced by Wigner negativity, which motivates us to now review the main properties of the Wigner function. For a quantum state described by the density matrix $\hat{\rho}$ on phase space $(x,p)$, the Wigner function is defined as the Wigner-Weyl transform of the density matrix~\cite{Wigner:1932}
\bea\label{eq:Wgendef}
    W(x, p) = \frac{1}{2\pi} \int_{-\infty}^{\infty} \dd u\, e^{- i p u} \left\langle x - \frac{u}{2} \bigg\vert \, \hat{\rho} \, \bigg\vert x + \frac{u}{2} \right\rangle \,.
\eea
Since the Wigner-Weyl transform is invertible, this provides a representation of the quantum state which is fully equivalent to the density matrix~\cite{2008AmJPh..76..937C}. One of the reasons why this representation is useful is that the expectation value of any operator $\hat{O}$ can be written as 
\bea
    \langle \hat{O}\rangle = \mathrm{Tr}(\hat{O}\hat{\rho}) = \int \dd x \dd p\, O(x,p) W(x,p)\, ,
\eea 
where $O(x,p)=(2\pi)^{-1}\int\dd u\langle x-u/2\vert \hat{O} \vert x+u/2\rangle$ is the Wigner-Weyl transform of $\hat{O}$. Since the Wigner function is real (given that $\hat{\rho}$ is Hermitian) and normalised to one ($\int \dd x \dd p W(x,p)=1$, since $\mathrm{Tr}(\hat{\rho})=1$), one may thus view $W$ as a quasi-distribution function against which observables can be computed. The precautionary prefix ``quasi'' is a reminder that the Wigner function is not necessarily everywhere positive. When it is positive, the state may be viewed as classical in the sense that it can be represented by a distribution function describing stochastic processes in phase space.\footnote{This led John Bell to argue in 1986 that states with positive Wigner functions cannot violate Bell inequalities~\cite{Bell:1986}. In 2005, Revzen showed that this statement is correct only if Bell operators are constructed out of proper spin operators (\ie~operators whose Wigner-Weyl transform takes value within the operator's spectrum), but that Bell inequality can otherwise be violated even if the Wigner distribution is positive definite~\cite{PhysRevA.71.022103, Revzen2006-REVTWF}. See~\cite{Martin:2019wta} for a historical perspective on this issue, and \cite{Martin:2016tbd} for explicit instances of Bell-inequality violations with positive Wigner functions and improper pseudo-spin operators.} By contrast, regions of negativity in the Wigner function mark a departure from classical behaviour~\cite{Kenfack:2004}.

Wigner negativity provides a phase-space signature of the interferences arising from quantum superpositions. As an illustration, consider the quantum cat state 
\bea
    \vert\psi\rangle = \frac{1}{\sqrt{2}}\left(\vert\psi_1\rangle+\vert\psi_2\rangle \right) \,,
\eea 
where $\psi_1$ and $\psi_2$ are each Gaussian wavefunctions. The Wigner function of such a state is given by 
\bea
    W=\frac{1}{2}\left(W_1+W_2\right)+W_{\mathrm{int}} \,,
\eea 
where $W_1$ and $W_2$ are the Wigner functions of $\psi_1$ and $\psi_2$, respectively. While these are Gaussian, and hence everywhere positive, the cross-term
\bea 
    W_{\mathrm{int}}(x,p) \propto \cos\left(\Delta\bar{x} p + \Delta\bar{p} x\right) \,,
\eea
features interference fringes with alternating positive and negative regions, which encode phase coherence. In this expression, $\Delta\bar{x} = \langle\psi_1\vert\hat{x}\vert\psi_1\rangle - \langle\psi_2\vert\hat{x}\vert\psi_2\rangle $ and $\Delta\bar{p} = \langle\psi_1\vert\hat{p}\vert\psi_1\rangle - \langle\psi_2\vert\hat{p}\vert\psi_2\rangle $ denote the relative differences in mean position and momentum, respectively. If the state decoheres, then the density matrix reduces to a statistical mixture $\hat{\rho} \rightarrow (\vert\psi_1\rangle \langle\psi_1\vert + \vert\psi_2\rangle \langle\psi_2\vert )/2$, for which $W_{\mathrm{int}}=0$ and the Wigner function becomes strictly positive. For these reasons, Wigner negativity, defined as 
\bea\label{eq:negativitydef}
    \mathcal{N} = \frac{1}{2}\left[\int \dd x \dd p \vert W(x,p) \vert - 1\right] \,,
\eea 
quantifies coherence between classically distinct configurations and is commonly used to monitor interference. In the context of resource theories, Wigner negativity is also a resource for quantum advantage~\cite{Albarelli:2018ujl}.

The main motivation for this work comes from Hudson's theorem, which states that \emph{a pure state has a non-negative Wigner function if and only if it is Gaussian}~\cite{HUDSON1974249, 1983JMP....24...97S}. As a contrapositive, any non-Gaussian pure state exhibits Wigner negativity. This is where primordial fluctuations rejoin the discussion: at leading order in perturbation theory, they are described by a quadratic Hamiltonian, and so provided they are initiated in a Gaussian state (as in the Bunch-Davies vacuum), they remain Gaussian and the Wigner function remains positive. General Relativity being non-linear, at higher order in perturbation theory the Hamiltonian receives cubic and higher order corrections. These terms render the state non-Gaussian, and hence the Wigner function non-positive.

In this work, we quantify the emergence of Wigner negativity in primordial fluctuations generated by non-linear dynamics during inflation. We develop a non-perturbative treatment of the wavefunction and its associated Wigner function in order to track phase space interference beyond the Gaussian approximation. We focus throughout on the closed-system (unitary) evolution of a pure state, leaving the inclusion of environmental decoherence to future work.

The remainder of this article is structured as follows: In Sec.~\ref{sec:formalism} we introduce our formalism, including the EFT of inflation as well as a homogeneous matching procedure inspired by the separate-universe approach. In Sec.~\ref{sec:constant roll}, we specialise to a general constant-roll background and solve for the wavefunction and Wigner function for the Goldstone of the EFT of inflation (which is closely related to the curvature perturbation). In Sec.~\ref{sec:results}, we present our results for the time evolution, parameter dependencies, and negativity of the Wigner function in an ultra-slow roll background. Finally, we conclude in Sec.~\ref{sec:conclusions} by discussing the implications of our results, as well as outlining a number of future directions.

\section{General formalism}\label{sec:formalism}

\subsection{EFT of inflation}

In order to capture non-Gaussian features in the quantum state of primordial perturbations, we need to go beyond linear cosmological perturbation theory, since in linear theory the Hamiltonian is quadratic and starting from the Bunch-Davis vacuum the Wigner function always remains Gaussian. One possibility is to work at higher order in cosmological perturbation theory, \ie~include cubic or higher terms in the action, since those lead to non-Gaussianities. However, to remain in the validity domain of the perturbative expansion, only small fractional modifications to the Wigner function can be studied with such an approach, and this excludes sign changes.\footnote{The notion that incorporating non-Gaussianities in a perturbative way can never yield large Wigner negativity is supported by the conclusions of~\cite{Haque:2025pav}.} This is why, in order to reveal potential non-positivity of the Wigner function, one needs to resort to non-perturbative techniques. 

One such non-perturbative scheme is the Effective Field Theory (EFT) of inflation~\cite{Cheung:2007}, which we briefly review. For simplicity, we consider a model of single-field inflation where the background inflaton field evolves according to some homogeneous solution $\phi_{\mathrm{b}}(t)$. This background defines a preferred time-slicing and spontaneously breaks time translation invariance. As a result, inflaton perturbations $\delta \phi(t,\mathbf{x})$ transform non-linearly with respect to time diffeomorphisms,
\begin{equation}
    t \rightarrow t + \xi^0(t,\mathbf{x}) \,, \,\,\, \delta \phi(t,\mathbf{x}) \rightarrow \delta \phi(t,\mathbf{x}) + \dot{\phi}_{\mathrm{b}}(t) \xi^0(t,\mathbf{x}) \,.
\end{equation}
To construct an action for perturbations in this background, we first go to unitary gauge, where inflaton fluctuations vanish $\delta \phi = 0$ and all perturbations are encoded in the metric. In this gauge, $\delta \phi$ can be interpreted as having been eaten by the graviton, in analogy to how Goldstones are eaten by vector bosons in unitary gauge of spontaneously-broken gauge theories. We then write down all operators which respect the remaining unbroken spatial diffeomorphism, and finally restore full diffeomorphism invariance by promoting the parameter $\xi^0(x)$ to a field $\chi(x)$, which non-linearly realises the time-translation symmetry,
\begin{equation}
\label{eq:chi:xi0}
    \chi(x) \rightarrow \chi(x) - \xi^0(x) \,.
\end{equation}

The field $\chi(x)$ should be interpreted as the Goldstone mode, which also describes scalar perturbations around the Friedmann-Lema\^itre-Robertson-Walker (FLRW) solution. The main advantage of re-introducing the Goldstone boson (rather than having it eaten) is that it makes the high-energy behaviour manifest. In particular, at high energies the mixing between the Goldstone and the graviton polarizations vanishes and so the two sectors decouple. Since mixing terms are suppressed by powers of the first slow-roll parameter $\epsilon_1$ or $1/\Mp^2$, where $\Mp$ is the reduced Planck mass, this scale can generally be quite low.\footnote{Formally, the decoupling limit corresponds to taking $\Mp^2 \rightarrow \infty$ and $\epsilon_1 \rightarrow 0$ such that $\Mp^2 \epsilon_1 = \text{constant}$.} Above it, we can study the physics of the Goldstone whilst neglecting metric fluctuations~\cite{Baumann:2011,Green:2024}.

In this so-called ``decoupling'' limit, a Hamiltonian that is valid to all orders in perturbation theory can be derived as follows~\cite{Firouzjahi:2025}. Consider the action for the matter and gravitational sectors of an FLRW quasi-de Sitter inflationary background $S = S_{\rm matt} + S_{\rm grav}$, where $S_{\rm grav}$ is the usual Einstein-Hilbert action. In the decoupling limit, we neglect gravitational backreaction and treat the metric as unperturbed. In practice, this amounts to taking the ADM lapse and shift to be non-dynamical and trivial, $N = 1$ and $N^i = 0$. All perturbations are then encoded in the matter sector, and we can focus on the decoupled action for $\chi$, which reads\footnote{This expression also assumes a canonically normalised and minimally coupled inflaton field with sound speed $\cs = 1$.}~\cite{Cheung:2007}
\begin{equation}\label{eq:originalaction}
    S = - \Mp^2 \int\! \dd^4 x \, a^3(t) \left\lbrace 3H^2(t+\chi) + 2 \left( 1 + \dot{\chi} \right) \dot{H}(t + \chi) + \dot{H}(t + \chi) \left[ \dot{\chi}^2 - \frac{1}{a^2(t)} (\partial \chi)^2 \right] \right\rbrace \, .
\end{equation}
In this expression, time is labeled by cosmic time $t$, and $H$ denotes the Hubble expansion rate, evaluated on the background solution. Notice that what appears in the argument of $H$ is not $t$, but rather $t + \chi$, since this is the combination that transforms linearly under time translations. Notice also that all the interactions are encoded in the $\chi$-dependence of the background dynamics. 

This action can be simplified considerably by working to leading order in $\epsilon_1\equiv -\dot{H}/H^2$, the so-called ``decoupling limit'' mentioned above. First, we can rewrite
\begin{equation}
    \dot{H}(t+\chi) = - H^2(t+\chi) \epsilon_1(t+\chi) = - H^2(t) \epsilon_1(t+\chi) + \mathcal{O}(\epsilon_1^2) \,.
\end{equation}
Additionally, using that $H(t+\chi) - H(t)=\mathcal{O}(\epsilon_1)$, we can write
\begin{equation}
    H^2(t+\chi) = \left\lbrace H(t) + \left[H(t+\chi) - H(t)\right]\right\rbrace^2 = 2 H(t) H(t+\chi) - H^2(t) + \mathcal{O}(\epsilon_1^2) \,.
\end{equation}
The first two terms in Eq.~\eqref{eq:originalaction} can then be combined into a total derivative term, and the action can be rewritten as
\begin{equation}\label{eq:S:before:canonical:transform}
    S = \Mp^2 \int \dd^4 x \, a^3(t) H^2(t) \epsilon_1(t+\chi) \left[ \dot{\chi}^2 - \frac{1}{a^2(t)}(\partial \chi)^2 \right] - \Mp^2 \int \dd^4 x \, \frac{\dd}{\dd t} \left[ 2 a^3(t) H(t+\chi) \right] \,,
\end{equation}
where we have dropped terms $\mathcal{O}(\epsilon_1^2)$ and higher, as well as a field-independent constant. The total-derivative term can be absorbed by performing the canonical transformation $\chi=\tilde{\chi}$ and $\pi = \tilde{\pi}+2 \Mp^2 a^3(t) H^2(t) \epsilon_1(t+\tilde{\chi})$, as shown\footnote{As numerous canonical transformations are employed in this work, we present a full derivation of the first in this appendix and omit analogous steps in subsequent cases.} in Appendix~\ref{app:boundaryterms}. We will use these transformed variables but drop the tildes for notational convenience. The result is the dramatically simplified action
\begin{equation}\label{eq:action}
    S =  \Mp^2 \int \dd^4 x \, a^3(t) H^2(t) \epsilon_1(t+\chi) \left[ \dot{\chi}^2 - \frac{1}{a^2(t)} (\partial \chi)^2 \right] \,,
\end{equation}
which matches the result of~\cite{Creminelli:2024cge, Firouzjahi:2025}. Notice that $\chi$ no longer appears in the argument of $H$, but does appear in the argument of the first slow-roll parameter, $\epsilon_1(t+\chi)$. 

\subsection{Relating the Goldstone boson to the curvature perturbation}\label{sec:Goldstone}

The Goldstone of broken time diffeomorphisms, $\chi$, can be related to the comoving curvature perturbation $\mathcal{R}$ as follows. In comoving gauge, the Goldstone vanishes and the comoving curvature perturbation enters in the spatial components of the metric $h_{ij}$ as
\begin{equation}
    \chi(t) = 0 \,, \,\,\, h_{ij}(t) = a^2(t) e^{2 \mathcal{R}} \delta_{ij} \,.
\end{equation}
Meanwhile in spatially flat gauge, the curvature perturbation vanishes $\mathcal{R} = 0$ but $\chi$ is finite, 
\begin{equation}
    \chi(\tilde{t}) \neq 0 \,, \,\,\, h_{ij}(\tilde{t}) = a(\tilde{t})^2 \delta_{ij} \,.
\end{equation}
These two coordinate systems are related by a time diffeomorphism, $\tilde{t} = t + \xi_0$. By equating the metrics, we can solve for the curvature perturbation as~\cite{Maldacena:2002, Cheung:2007sv, Behbahani:2011it}
\begin{equation}
    \mathcal{R}(t,\mathbf{x}) = \ln \left[ \frac{a(t + \xi_0)}{a(t)} \right] = \int_t^{t + \xi_0(t,\mathbf{x})} \dd t' \, H(t') \,.
\end{equation}
Formally, this can be solved as
\begin{equation}\label{eq:Rpert}
    \mathcal{R} = \sum_{n=1}^\infty \frac{1}{n!} H^{(n-1)}(t) \xi_0^n \,,
\end{equation}
where $H^{(n)} = \frac{\dd^n}{\dd t^n} H$, and in the decoupling limit this reduces to $\mathcal{R}=H \xi_0$.

Moreover, under a time diffeomorphism, recall that the Goldstone shifts according to Eq.~(\ref{eq:chi:xi0}). In particular in going from flat gauge to comoving gauge, we have $\chi(\tilde{t}) \rightarrow \chi(\tilde{t}) + \xi_0$. By equating this to the value of the Goldstone in comoving gauge $\chi(t) = 0$, we obtain the equation
\begin{equation}\label{eq:chi:t:xi0:implicit:eq}
    \chi(t + \xi_0) + \xi_0 = 0 \,,
\end{equation}
which implicitly relates $\xi_0$ to $\chi$. A formal solution to this equation is given by
\bea\label{eq:chi:t:xi0:implicit:sol} 
    \xi_0(t) = \sum_{n=1}^\infty \frac{(-1)^n}{n!} \left[\chi^n(t) \right]^{(n-1)} \,,
\eea
which follows from an application of the Lagrange inversion theorem\footnote{The Lagrange inversion theorem~\cite{Lagrange:1770,Gessel:2016} allows one to obtain the formal series expansion of an inverse function. Given an implicit equation of the form $y = t + a f(y)$, one has the formal series solution
\bea\label{eq:Lagrangeinversion}
    y(t) = t + \sum_{n=1}^\infty \frac{a^n}{n!} \frac{\partial^{n-1}}{\partial t^{n-1}} \left[ f(t)^n \right] \,.
\eea
In order to apply this to solve Eq.~(\ref{eq:chi:t:xi0:implicit:eq}), we perform the change of variables $y(t) = t + \xi_0(t)$, which brings the equation into the form $y = t - \chi(y)$. One can then use Eq.~(\ref{eq:Lagrangeinversion}) with $a = -1$ to obtain the solution in Eq.~(\ref{eq:chi:t:xi0:implicit:sol}).}. Combining this result with Eq.~(\ref{eq:Rpert}) and using the general Leibniz rule combined with the Cauchy product rule finally leads to the following non-linear relationship between $\mathcal{R}$ and $\chi$, 
\bea\label{eq:nonlinearR}
    \mathcal{R} = \sum_{n=1}^\infty \frac{(-1)^n}{n!} (H \chi^n)^{(n-1)} \,.
\eea
If the curvature perturbation is conserved, then in the decoupling limit $\xi^0=\mathcal{R}/H$ is conserved too, and Eq.~(\ref{eq:chi:t:xi0:implicit:eq}) implies that $\chi$ is conserved as well, and
\begin{equation}\label{eq:linearchiR}
    \mathcal{R} = - H \chi \,.
\end{equation}
Otherwise, the relation between $\chi$ and $\mathcal{R}$ remains non-linear even in the decoupling limit. Note, however, that since the Lagrangian in Eq.~(\ref{eq:action}) is proportional to $\partial_\mu\chi\partial^\mu\chi$, $\chi = \mathrm{constant}$ always provides a solution to the complete non-linear equations of motion. This is why $\chi$ plays no less fundamental role than $\mathcal{R}$, and in what follows we will study its quantum state rather than the one of $\mathcal{R}$.

\subsection{Canonical normalisation}\label{sec:canonical}

Before solving for the quantum state of the Goldstone boson $\chi$, let us see how the action in Eq.~\eqref{eq:action} can be rewritten in canonical form, \ie~with a canonical kinetic term. Focusing of the homogeneous mode $\chi_0(t)$, the Lagrangian is of the form
\bea\label{eq:zero:mode:Lagrangian}
    L\left(\chi_0,\dot{\chi}_0, t \right) = f(t,\chi_0)\dot{\chi}_0^2 \,, \quad \text{with} \quad f(t,\chi_0)= \mathcal{V} \Mp^2 a^3(t) H^2(t) \epsilon_1(t+\chi_0) \,,
\eea
where $\mathcal{V} = \int \dd ^3 x$ is the comoving volume coming from the spatial integral in Eq.~(\ref{eq:action}). Let us introduce the new variable
\bea\label{eq:x:from:chi}
    X(t,\chi_0) = & \int_{\chi_{\mathrm{ref}}}^{\chi_0} \dd\tilde{\chi} \sqrt{2 f(t, \tilde{\chi})}\, ,
\eea  
where $\chi_{\mathrm{ref}}$ is an arbitrary reference value. One has
\bea
    \dot{X} = \sqrt{2 f(t,\chi_0)} \dot{\chi}_0 +  \int_{\chi_{\mathrm{ref}}}^{\chi_0} \dd\tilde{\chi} \frac{\partial_t f(t, \tilde{\chi})}{\sqrt{2 f(t, \tilde{\chi})}} \,,
\eea
such that the Lagrangian becomes
\bea\label{eq:Lwithcrossterm}
    L (X,\dot{X}, t)= \frac{1}{2}\left[ \dot{X}-  A\left(t, X\right) \right]^2 \,,
\eea
where we have defined
\bea\label{eq:A:def}
    A\left(t, X\right) = \int_{\chi_{\mathrm{ref}}}^{\chi_0(t,X)} \dd\tilde{\chi} \frac{\partial_t f(t, \tilde{\chi})}{\sqrt{2 f(t, \tilde{\chi})} } \,,
\eea
and $\chi_0(t,X)$ denotes the inverse of $X(t,\chi_0)$. The term in $L$ which is linear in $\dot{X}$ can be removed by subtracting a total time derivative. We introduce
\bea\label{eq:FArelation}
    F\left(t, X \right)= \int_{X_{\mathrm{ref}}}^X \!\! \dd\tilde{X} \, A(t, \tilde{X}) \,, 
\eea
the total time derivative of which is
\bea
    \frac{\dd}{\dd t} F\left(t, X \right) = A\left(t, X \right) \dot{X} + \int_{X_{\mathrm{ref}}}^X \!\! \dd\tilde{X} \, \frac{\partial}{\partial t}A(t,\tilde{X}) \,.
\eea
By substituting the term $\dot{X}A(t,X)$ appearing in the Lagrangian with the expression above, one obtains 
\bea\label{eq:Lwithtotderiv}
    L (X,\dot{X}, t) = \frac{1}{2} \dot{X}^2 - V(t,X) - \frac{\dd}{\dd t} F\left(t,X \right) \,,
\eea
where we have defined the potential
\bea\label{eq:V:def}
    V\left(t, X\right) =& - \frac{1}{2} A\left(t, X\right)^2 -\int_{X_{\mathrm{ref}}}^X \!\! \dd\tilde{X} \, \frac{\partial}{\partial t} A(t, \tilde{X}) \, .
\eea
The total derivative term can be removed by performing the canonical transformation $X = \tilde{X}$, $\pi_X = \tilde{\pi}_X - A(t,X)$, as we demonstrate in Appendix~\ref{app:boundaryterms}. The Lagrangian is thus cast in a form where the kinetic term is canonical, 
\bea\label{eq:Lagrangian:canonical}
    L (X,\dot{X}, t) = \frac{1}{2} \dot{X}^2 - V(t,X) \,,
\eea
although it now contains a time-dependent self-interaction potential $V(t, X)$.

\subsection{Matching prescription}\label{sec:matchingprescription}

The canonical Lagrangian~\eqref{eq:Lagrangian:canonical} describes the dynamics of the homogeneous mode\footnote{The ``homogeneous mode'' $\chi_0(t)$ should be understood as a proxy for some appropriately coarse-grained long-wavelength field, rather than as the literal global homogeneous field configuration. Equivalently, $\chi_0(t)$ can be interpreted as the leading-order term in the gradient expansion.} $\chi_0(t)$; hence it only applies to cosmological fluctuations whose spatial extent $\lambda$ far exceeds the Hubble radius, $\lambda<\lambda_\sigma \equiv (\sigma H)^{-1}$, where $\sigma \ll 1$. At near- and sub-Hubble scales, keeping track of the quantum state of $\chi$ is more involved since it requires solving the Schr\"odinger equation for a wavefunctional, rather than a wavefunction. This is why, in this work, we will adopt the same strategy as in the $\delta N$ formalism~\cite{Starobinsky:1982, Starobinsky:1985, Sasaki:1995, Sasaki:1998, Lyth:2004} or the stochastic-$\delta N$ formalism~\cite{Fujita:2013, Fujita:2014, Vennin:2015}, where perturbation theory is employed within $\lambda_\sigma$, and above $\lambda_\sigma$ one uses the separate-universe approach~\cite{Salopek:1990, Sasaki:1995, Wands:2000, Lyth:2003, Rigopoulos:2003, Lyth:2005}, according to which the universe can be described as a collection of independent FLRW patches. This implies that the results derived below only account for non-linearities at large scales, which are nonetheless expected to be the most relevant ones since, inside the Hubble radius, cosmological fluctuations are not subject to the parametric amplification driven by the background expansion, and thus remain close to their vacuum Gaussian state. 

The above scheme implies that a matching procedure is enforced at the scale $\lambda_\sigma$, in order to set initial conditions for the homogeneous-mode Schr\"odinger equation. In practice, we consider fluctuations with a physical wavelength $\lambda$ (or equivalently, with comoving wavenumber $k = a/\lambda$), and we denote by $t_\star$ the time at which $k = \sigma a_\star H$. Since linear perturbation theory is employed to describe the phase $t<t_\star$, at time $t_\star$ the wavefunction of $X$ is still Gaussian,\footnote{More generally, the wavefunction is of the form $\psi_X = N e^{- \alpha (X - \bar{X})^2}$, where a non-zero $\bar{X}$ allows for a non-vanishing $\langle \hat{X} \rangle$ and $\langle \hat{\pi}_X \rangle$. For our choice of reference point $\chi_{\rm ref} = 0$, $\bar{X} = 0$ at linear order.\label{foonote:Xbar}}
\begin{equation}\label{eq:psi:X:tleqtstar}
    \psi_X(t, X) = N(t) e^{ - \alpha(t) X^2}\quad\text{for}\quad t\leq t_\star \,,
\end{equation}
where $\vert N\vert =(2\alphaR/\pi)^{1/4}$ for the state to be properly normalized. The wavefunction being Gaussian, the parameter $\alpha=\alphaR + i \alphaI$ is entirely determined by the quadratic moments of $X$ and its conjugate momentum $\pi_X$, 
\bea 
\label{eq:<X2>}
    \langle \hat{X}^2 \rangle = & \frac{1}{4\alphaR} \,,\\
    \langle \hat{\pi}_X^2 \rangle = & \frac{|\alpha|^2}{\alphaR} \,,\\
    \langle \hat{X} \hat{\pi}_X \rangle =  & \frac{i}{2} -\frac{\alphaI}{2\alphaR} \,,
\eea 
where we have used the representation $\hat{\pi}_X = -i \partial_X$ for the conjugate momentum. The above can be readily inverted to find
\bea\label{eq:alpha:moments}
    \alpha = -\frac{i}{2}\frac{\langle \hat{X} \hat{\pi}_X \rangle}{\langle \hat{X}^2 \rangle}\, .
\eea 

Meanwhile, from linear perturbation theory one can compute the power spectra $\mathcal{P}_{\chi\chi}(t,k)$, $\mathcal{P}_{\pi\pi}(t,k)$, and $\mathcal{P}_{\chi\pi}(t,k)$, where in general we define the power spectrum $\mathcal{P}_{gh}$ for the fields $\hat{g}$ and $\hat{h}$ as
\bea\label{eq:Pkgendef}
    \frac{k^3}{2\pi^2} \langle \hat{g}_{\mathbf{k}} \hat{h}^\dagger_{\mathbf{k}'} \rangle =  \mathcal{P}_{gh}(k) \delta ^{(3)}(\mathbf{k}-\mathbf{k}') \,.
\eea
In real space, $\langle \hat{g}(\mathbf{x}) \hat{h}(\mathbf{x})\rangle = \bar{g} \bar{h} + \int \dd\ln k \, \mathcal{P}_{hg}(k)$, where $\bar{g} \equiv \langle \hat{g}(\mathbf{x}) \rangle$ is the (possibly non-vanishing) one-point function. As described above, our goal is to describe fluctuations at a given comoving scale $k$, employing perturbation theory before a certain matching time and the homogeneous-mode description afterwards. The matching can thus be performed by identifying the two-point functions of the quantised homogeneous field operators $\hat{\chi}_0$ and $\hat{\pi}_0$ with their linear power spectra as\footnote{We emphasise that $\bar{\chi}_0 \equiv \langle \hat{\chi}_0 \rangle$ and $\bar{\pi}_0 \equiv \langle \hat{\pi}_0 \rangle$ are quantum expectation values, not spatial averages.}
\bea\label{eq:gh:Deltalnk}
    \langle \hat{\chi}_0^2 \rangle & \to \bar{\chi}_0^2  + \mu_1 \mu_2 \mathcal{P}_{\chi\chi}(t_\star,k)\,, \\
    \langle \hat{\chi}_0 \hat{\pi}_0 \rangle & \to \bar{\chi}_0 \bar{\pi}_0  + \mu_1 \mathcal{P}_{\chi\pi}(t_\star,k)\,, \\
    \langle \hat{\pi}_0^2 \rangle & \to \bar{\pi}_0^2  + \frac{\mu_1}{\mu_2} \mathcal{P}_{\pi\pi}(t_\star,k)\, .
\eea 
Two parameters appear in this identification, which play different roles~\cite{Martin:2021xml}. The first, $\mu_1$, must be chosen such that the homogeneous fields $\chi_0$ and $\pi_0$ are canonically normalised, \ie~$\langle [\hat{\chi}_0,\hat{\pi}_0] \rangle = 2 i \mathrm{Im}[ \langle \hat{\chi}_0 \hat{\pi}_0 \rangle ] =i$. Since the Fourier modes $\hat{\chi}_{\mathbf{k}}$ and $\hat{\pi}_{\mathbf{k}}$ are canonically normalised, $[ \hat{\chi}_{\mathbf{k}}, \hat{\pi}^\dagger_{\mathbf{k}'}] = i \delta^{(3)}(\mathbf{k}-\mathbf{k}')$, one has $\mathrm{Im}[\mathcal{P}_{\chi\pi}(k)]=k^3/4\pi^2$, which leads to
\bea\label{eq:mu:def}
    \mu_1 = \frac{2\pi^2}{k^3} \,.
\eea
This is also a natural choice for the comoving volume factor $\mathcal{V}$, since the homogeneous mode is extracted by coarse graining the field over an effective volume $\mathcal{V} \sim k^{-3}$. Though the precise relation between $\mu_1 = 2\pi^2/k^3$ and $\mathcal{V}$ will depend on the window function chosen for the coarse graining (see~\cite{Martin:2021xml}), they should be proportional up to some $\mathcal{O}(1)$ geometrical factor. Since the choice of this factor will not affect any physical observables, we set $\mathcal{V} = \mu_1$ without loss of generality. The second parameter, $\mu_2$, sets the units in which $\chi_0$ and $\pi_0$ are measured. Because of the form of the Lagrangian~\eqref{eq:zero:mode:Lagrangian}, $\chi_0$ has inverse-mass dimension,\footnote{From Eqs.~(\ref{eq:action}) and (\ref{eq:zero:mode:Lagrangian}), we see that $\chi_0(t)$ and $\chi(t,\mathbf{x})$ have the same mass dimension. The same cannot be true for $\pi_0(t)$ and $\pi(t,\mathbf{x})$, since $[\chi_0,\pi_0]=i$ is dimensionless while $[\chi(\mathbf{x}),\pi(\mathbf{x}')]=i\delta(\mathbf{x}-\mathbf{x}')$ has the dimension of an inverse comoving volume. Another way to see why $\pi_0$ and $\pi$ do not have the same dimension is to note that $\pi_0$ is defined with respect to the Lagrangian as $\pi_0 = \partial L/\partial \dot{\chi}_0$ whereas $\pi$ is defined with respect to the Lagrangian \textit{density} $\mathcal{L}$, $\pi = \partial \mathcal{L}/\partial \dot{\chi}$. Explicitly, $[\chi_0] = [\chi] = M^{-1}$ while $[\pi_0] = M$ and $[\pi] = M^4$.} and hence $\mu_2$ must have the dimension of an inverse comoving volume. A natural choice is then
\bea\label{eq:mu2:def}
    \mu_2 = \frac{1}{\mathcal{V}} = \frac{1}{\mu_1} \,.
\eea 
Ultimately, since changing $\mu_2$ simply amounts to performing a linear canonical transformation, which does not affect the properties discussed in Sec.~\ref{sec:results} (in particular, the Wigner negativity), we adopt Eq.~(\ref{eq:mu2:def}) without loss of generality.

Moreover, linearising Eq.~(\ref{eq:x:from:chi}) gives $X=\sqrt{2f(t,0)}\, \chi_0$ provided we set $\chi_{\mathrm{ref}}=0$. This will be our convention going forward since, as discussed in footnote~\ref{foonote:Xbar}, this is the unique choice for which $\langle \hat{X} \rangle$ vanishes at linear order, greatly simplifying the matching conditions. From Eq.~(\ref{eq:Lagrangian:canonical}), one has $\pi_X = \dot{X}$, and from Eq.~(\ref{eq:zero:mode:Lagrangian}), $\pi_0=2f(t,0)\dot{\chi}_0$. The conjugate momenta are then related as $\pi_X=[\pi_0 + \dot{f}(t,0) \chi_0 ] / \sqrt{2f(t,0)}$ at linear order. Together with Eqs.~(\ref{eq:alpha:moments}) and \eqref{eq:gh:Deltalnk}, this leads to
\bea
\label{eq:alphaLpower:spectra:chi}
    \alpha_\star = -\frac{i}{4}\frac{\dot{f}(t_\star,0)}{f(t_\star,0)} -\frac{i}{4} \frac{\mathcal{V}}{f(t_\star,0)} \frac{\mathcal{P}_{\chi\pi}(t_\star,k)}{\mathcal{P}_{\chi\chi}(t_\star,k)} \,,
\eea 
where $\alpha_\star \equiv \alpha(t_\star)$. Introducing the second Hubble-flow parameter $\epsilon_2 \equiv \dot{\epsilon}_1/H\epsilon_1$, the first term can be evaluated by noting that $\dot{f}(t,0)/f(t,0) = (3 + \epsilon_2) H$. The second term can be re-written in terms of power spectra of the curvature perturbation, which can be more easily computed in cosmological perturbation theory. Linearising Eq.~(\ref{eq:nonlinearR}) to obtain $\mathcal{R} = - H \chi$ and also noting that the momentum conjugate to the full field $\chi$ in Eq.~(\ref{eq:action}) is $\pi \equiv \partial \mathcal{L}/\partial \dot{\chi} = 2f(t_\star,0)\dot{\chi}/\mathcal{V}$, we have $\mathcal{P}_{\mathcal{R}\mathcal{R}} = H^2 \mathcal{P}_{\chi\chi}$ and $\mathcal{P}_{\mathcal{R} \dot{\mathcal{R}}} = [H^2\mathcal{V}/2 f(t,0)] \mathcal{P}_{\chi \pi}$. Thus, Eq.~(\ref{eq:alphaLpower:spectra:chi}) becomes
\bea\label{eq:alphaLpower:spectra}
    \alpha_\star = -\frac{i}{4}\left(3+\epsilon_2\right)H - \frac{i}{2} \frac{\mathcal{P}_{\mathcal{R}\dot{\mathcal{R}}}(t_\star,k)}{\mathcal{P}_{\mathcal{R}\mathcal{R}}(t_\star,k)} \,,
\eea 
where conveniently one notes that the factor of $\mathcal{V}$ has cancelled out. 

In linear perturbation theory, the mode function for the comoving curvature perturbation satisfies 
\begin{equation}\label{eq:MS}
    \mathcal{R}_k'' + 2 \frac{z'}{z} \mathcal{R}_k' + k^2 \mathcal{R}_k = 0 \,,
\end{equation}
where primes denote derivatives with respect to conformal time $\eta = \int \dd t/a$ and $z^2 = 2 \Mp^2 a^2 \epsilon_1$ is the usual Mukhanov-Sasaki pump field, which governs the coupling of scalar perturbations to the background evolution. Upon solving this equation with appropriately chosen vacuum conditions, the power spectra in Eq.~(\ref{eq:alphaLpower:spectra}) can be evaluated according to $\mathcal{P}_{\mathcal{R}\dot{\mathcal{R}}}/\mathcal{P}_{\mathcal{R}{\mathcal{R}}}=\dot{\mathcal{R}}_k^*/\mathcal{R}_k^*$.

\section{Constant-roll inflation}\label{sec:constant roll}

Let us now show how the Wigner function of the Goldstone boson $\chi_0$ can be derived in the above framework. For concreteness, we consider a phase of inflation where the second Hubble-flow parameter $\epsilon_2 = \dot{\epsilon}_1/H \epsilon_1$ is constant. A constant value for $\epsilon_2$ leads to a very simple time dependence for the first Hubble-flow parameter $\epsilon_1$, which will allow us to carry out explicit calculations. This regime is often referred to as ``constant-roll'' inflation, and it encompasses the well-known cases of slow roll (SR) for $\epsilon_2 = 0$~\cite{Steinhardt:1984jj,Liddle:1992wi,Liddle:1994dx,Martin:2013tda} and ultra-slow roll (USR) for $\epsilon_2 = -6$~\cite{Kinney:2005,Motohashi:2012,Namjoo:2012,Dimopoulos:2017,Pattison:2018}.

\subsection{Wavefunction} \label{sec:Wavefunction}

If $\epsilon_2$ is constant, then in the decoupling limit one has
\bea\label{eq:eps1def}
    \epsilon_1(t) = \bar{\epsilon}_1 e^{\epsilon_2 H t}\, ,
\eea 
with $\bar{\epsilon}_1$ the value of $\epsilon_1$ at a reference time $t=0$, where we also set $a(0) = 1$. We take this reference to correspond to Hubble crossing, \ie $k = a H |_{t=0} = H$, such that $\bar{\epsilon}_1$ denotes the value of $\epsilon_1$ at horizon re-entry. The function $f$ introduced in Eq.~(\ref{eq:zero:mode:Lagrangian}) then reads 
\bea
    f(t,\chi_0) = \mathcal{V} \Mp^2 H^2 \bar{\epsilon}_1 e^{(3 + \epsilon_2) H t} e^{\epsilon_2 H \chi_0} \,,
\eea
where we have used that $a(t) = e^{Ht}$ in the decoupling limit. This function is separable in the sense that the $t$ and $\chi_0$ dependence in $\epsilon_1(t+\chi_0)$ factorises, which allows for dramatic simplifications in the procedure outlined in Sec.~\ref{sec:canonical}. In particular, Eqs.~(\ref{eq:x:from:chi}) and~\eqref{eq:A:def} lead to $A = \frac{1}{2} (3+\epsilon_2) H X$, and so the potential~\eqref{eq:V:def} reads 
\bea\label{eq:potential:X}
    V(t,X) = - \frac{1}{8} (3+\epsilon_2)^2 H^2 X^2 \,.
\eea 
That this potential is quadratic in $X$ implies that it admits Gaussian solutions, while non-Gaussian features are contained in the non-linear canonical transformation between $\chi_0$ and $X$. One also notes that this potential is the same for SR ($\epsilon_2 = 0$) and USR ($\epsilon_2 = -6$) backgrounds, which is a manifestation of the Wands duality~\cite{Wands:1998yp}. 

Though this potential is simple, time-independent, and quadratic, the change of variables relating $X$ and $\chi_0$ will introduce a time-dependent boundary condition which complicates solving the Schr{\" o}dinger equation. In anticipation of simplifying this boundary condition, we will take a slight variation on the change of variables presented in Sec.~\ref{sec:canonical}. It should be relatively straightforward to map the previous results onto this new variable. With this motivation in mind, we factorise out the time dependence in $f$ as
\bea
    f(t,\chi_0) = e^{(3+\epsilon_2) H t} g(\chi_0) \,, \quad g(\chi_0) = \mathcal{V} \Mp^2 H^2 \bar{\epsilon}_1 e^{\epsilon_2 H \chi_0} \,,
\eea
and introduce the change of variables
\bea\label{eq:Ycov}
    Y(\chi_0) = \int_0^{\chi_0} \dd \tilde{\chi} \, \sqrt{2 g(\tilde{\chi})} \,,
\eea
where as in Sec.~\ref{sec:matchingprescription} we have set $\chi_{\rm ref} = 0$. This is analogous to the definition of $X$ in Eq.~(\ref{eq:x:from:chi}), but crucially because $g(\chi_0)$ is time-independent, $\dot{Y} = \sqrt{2 g(\chi_0)} \dot{\chi}_0$ has no auxiliary $A$-piece. The Lagrangian is thus
\bea\label{eq:LagrangianY}
    L(Y, \dot{Y}, t) = \frac{1}{2} e^{(3 + \epsilon_2) H t} \, \dot{Y}^2 \,.
\eea
While the kinetic term is not canonical, and therefore $\pi_Y = e^{(3+\epsilon_2)Ht} \, \dot{Y}$ has a time-dependent\footnote{One can eliminate this time dependence by introducing a new time coordinate $\tau = - e^{- (3 + \epsilon_2) H t}/(3+\epsilon_2) H$, in terms of which the kinetic term becomes canonical. Because we have already introduced enough new variables, and because this method offers no real computational simplification, we will not do so here.} pre-factor, this brings no additional complications and can be accommodated by only a slight modification to the matching conditions.

Explicitly, the change of variables in Eq.~(\ref{eq:Ycov}) reads
\bea\label{eq:Y:chi0}
    Y = \frac{2 \kappa}{\epsilon_2 H} \left( e^{\frac{1}{2} \epsilon_2 H \chi_0} - 1 \right) \,, \quad \kappa = \sqrt{2 \mathcal{V} \Mp^2 H^2 \bar{\epsilon}_1} \,.
\eea
Notice that this relationship is generically non-linear, but reduces to linear for SR, $Y_{\rm SR} = \kappa \chi_0$. Note also that unless $\epsilon_2 = 0$, this is a compactifying change of variables, with the domain $\chi_0 \in (- \infty, \infty)$ mapped to 
\bea
    Y \in \begin{cases} (- \infty, Y_{\mathrm{b}}) & \epsilon_2 < 0 \,, \\
    (Y_{\mathrm{b}}, \infty) & \epsilon_2 > 0 \,, \end{cases}
\eea
where we have defined the finite boundary location
\bea
\label{eq:Yb:def}
    Y_{\rm b} = - \frac{2 \kappa}{\epsilon_2 H} \,.
\eea 
We focus on the former case $\epsilon_2 < 0$ without loss of generality. As $Y$ approaches $Y_{\mathrm{b}}$, $\chi_0$ approaches infinity, where its wavefunction must vanish for the quantum state to be normalisable. We therefore impose the Dirichlet boundary condition
\bea\label{eq:Dirichlet:boundary:condition}
    \psi_Y(Y_{\rm b}) = 0 \,.
\eea 
Note that the Gaussian ansatz used in the matching prescription of Sec.~\ref{sec:matchingprescription} does not satisfy this boundary condition. This mismatch is not problematic, however; the finite boundary is fundamentally a non-perturbative effect, whereas the matching condition is derived entirely within perturbation theory. In linear theory, using that $\langle \hat{Y}^2 \rangle = \kappa^2 \langle \hat{\chi}_0^2 \rangle \rightarrow (\kappa^2/H^2) \mathcal{P}_{\mathcal{R}\mathcal{R}}(t_\star,k)$, one finds
\bea
\label{eq:Y2:Yb:comp}
    \frac{\langle \hat{Y}^2 \rangle}{Y_{\rm b}^2} \bigg|_{t = t_\star} = \frac{\epsilon_2^2}{4} \mathcal{P}_{\mathcal{R} \mathcal{R}}(t_\star, k) \,.
\eea
Since the power spectrum must be small for perturbation theory to be valid prior to the matching time, the ratio of $\langle \hat{Y}^2 \rangle$ to $Y_{\rm b}^2$ must be small, ensuring that $Y_{\mathrm{b}}$ lies in the far tail of the wavefunction. The finite boundary is thus irrelevant to the linear-theory dynamics. Boundary effects and other non-perturbative features become important only afterwards, and it is precisely these phenomena that the super-Hubble homogeneous-mode description is intended to capture. 

Since the Hamiltonian $\mathsf{H} = \frac{1}{2} e^{-(3+\epsilon_2) H t} \, \pi_Y^2$ is quadratic, the Schr{\"o}dinger equation $i \partial_t \psi_Y = \hat{\mathsf{H}} \psi_Y$ admits Gaussian solutions. The simplest ansatz consisting of Gaussian wavepackets which satisfies the boundary condition~\eqref{eq:Dirichlet:boundary:condition} is
\bea\label{eq:Psi:Gaussian}
    \psi_Y(t,Y) = N(t) \left[ e^{- \beta(t) Y^2} - e^{- \beta(t) \left(Y - 2 Y_{\rm b} \right)^2} \right] \Theta\left(Y_{\rm b} - Y \right) \,,
\eea
where $\Theta$ is the Heaviside theta function restricting the domain to $Y \in (- \infty, Y_{\mathrm{b}})$. This approach, sometimes referred to as the ``method of images''~\cite{Kleber:1994}, is standard when describing Gaussian wavepackets incident on an infinite potential step in quantum mechanics. The second Gaussian can be interpreted as a reflection of the first wavepacket when it reaches the boundary. Since the ansatz~\eqref{eq:Psi:Gaussian} consists of a linear superposition of two Gaussians, it will be a solution provided each individual Gaussian satisfies the Schr{\"o}dinger equation.

Substituting Eq.~(\ref{eq:Psi:Gaussian}) into the Schr{\" o}dinger equation yields the following differential equation for the complex parameter $\beta$,
\bea\label{eq:betaeq}
    \dot{\beta} = - 2 i e^{- (3 + \epsilon_2) H t} \beta^2 \,,
\eea
or equivalently for the real $\betaR \equiv \text{Re}[\beta]$ and imaginary $\betaI \equiv \text{Im}[\beta]$ parts
\bea\label{eq:betaReIm}
    \dot{\beta}_{\rm R} = 4 e^{- (3 + \epsilon_2) H t} \betaR \betaI \,, \quad \dot{\beta}_{\rm I} = - 2 e^{-(3+\epsilon_2) Ht} (\betaR^2 - \betaI^2) \,.
\eea
Eq.~(\ref{eq:betaeq}) has the form of a Riccati equation, and can be solved by letting $\beta = - i e^{(3+\epsilon_2) Ht} \dot{\gamma}/2 \gamma$. The equation for $\gamma$ then takes the form $\ddot{\gamma} + (3+\epsilon_2) H \dot{\gamma} = 0$, which is the classical Euler-Lagrange equation corresponding to the Lagrangian of Eq.~(\ref{eq:LagrangianY}). The general solution is $\gamma = c_1 + c_2 e^{-(3+\epsilon_2) H (t-t_\star)}$, and using the initial condition $\beta(t_\star) \equiv \beta_\star$, one finds
\bea\label{eq:beta}
    \beta(t) = \frac{\beta_\star}{1 - \frac{2 i \beta_\star}{(3+\epsilon_2) H} \left[ e^{- (3 + \epsilon_2) H t} - e^{- (3 + \epsilon_2) H t_\star} \right]} \,.
\eea
The explicit form of $\beta_\star$ will be derived in Sec.~\ref{sec:constant:eps2:matching:conditions}. 

Upon substituting Eq.~(\ref{eq:Psi:Gaussian}) into the Schr{\"o}dinger equation, one also obtains the differential equation for the norm
\bea\label{eq:normeq}
    \frac{\dot{N}}{N} = - i e^{-(3+\epsilon_2) H t} \beta \,.
\eea
Meanwhile, in order for the wavefunction to be properly normalised, one must have
\bea\label{eq:normalisation}
    |N(t)| = \left( \frac{2 \betaR}{\pi} \right)^{1/4} \left( 1 - e^{-2 \frac{|\beta|^2}{\betaR} Y_{\rm b}^2} \right)^{-1/2} \,,
\eea
where $\betaR$ is required to be positive. One can verify that this solution also satisfies Eq.~(\ref{eq:normeq}) since $|\dot{N}|/|N| = \dot{\beta}_{\rm R}/4\betaR = e^{-(3+\epsilon_2) H t} \betaI$, where we have used the equation of motion for $\betaR$ as well as the fact that $|\beta|^2/\betaR$ is a constant in time, which may be seen by differentiating $\partial_t (|\beta|^2/\betaR)$ and using the equations of motion in Eq.~(\ref{eq:betaReIm}).

We can now transform back to the original variables $\chi_0$ and $\pi_0$ as follows. The relationship between $Y$ and $\chi_0$ is given in Eq.~(\ref{eq:Y:chi0}), while from Eq.~(\ref{eq:LagrangianY}) the momentum conjugate to $Y$ is $\pi_Y = e^{(3+\epsilon_2) H t} \dot{Y}$ and from Eq.~(\ref{eq:zero:mode:Lagrangian}) the momentum conjugate to $\chi_0$ is $\pi_0 = \kappa^2 e^{(3+\epsilon_2) H t} e^{\epsilon_2 H \chi_0} \dot{\chi}_0$.\footnote{For consistency with Sec.~(\ref{sec:matchingprescription}), we use here the conjugate momentum corresponding to the action in Eq.~(\ref{eq:action}) rather than that in Eq.~(\ref{eq:S:before:canonical:transform}). The canonical transformation that eliminates the total derivative term from the action~\eqref{eq:action} amounts to shifting the conjugate momentum so as to eliminate its 1-point function. Working from this action then has the nice property that the Wigner function will be centred in phase space.\label{footnote:momentum:shift}} Thus, we have
\bea\label{eq:chi0_pi0:Y_piY}
    Y = \underbrace{\,\frac{2\kappa}{\epsilon_2 H} \left( e^{\frac{1}{2} \epsilon_2 H \chi_0} - 1 \right)}_{h(\chi_0)} \,, \quad \pi_Y = \underbrace{\, \frac{1}{\kappa} e^{-\frac{1}{2} \epsilon_2 H \chi_0}\, \pi_0 \,}_{\pi_0/h'(\chi_0)} \,,
\eea
where $'$ denotes a derivative with respect to $\chi_0$. This transformation is of the form $Y = h(\chi_0)$ and $\pi_Y = \pi_0/h'(\chi_0)$, which manifestly preserves the Poisson-bracket structure. At the quantum mechanical level, the expression for $\pi_Y$ should be Weyl-ordered according to $\hat{\pi}_Y = \frac{1}{2}\left[\frac{1}{h'(\hat{\chi}_0)} \hat{\pi}_0 + \hat{\pi}_0 \frac{1}{h'(\hat{\chi}_0)}\right]$ in order for $\pi_Y$ to be Hermitian. Using that $\hat{\pi}_0 = - i \frac{\partial}{\partial \chi_0}$ in the position representation, the Weyl-ordered operator can equivalently be written as $\hat{\pi}_Y = \frac{1}{h'(\hat{\chi}_0)} \hat{\pi}_0 + \frac{i}{2} \frac{h''(\hat{\chi}_0)}{h'(\hat{\chi}_0)^2}$. At the operator level, the transformation then reads
\bea
    \hat{Y} = h(\hat{\chi}_0) \,, \quad \hat{\pi}_Y = \frac{1}{h'(\hat{\chi}_0)} \hat{\pi}_0 + \frac{i}{2} \frac{h''(\hat{\chi}_0)}{h'(\hat{\chi}_0)^2}  \,,
\eea
and the wavefunction transforms under it according to~\cite{Anderson:1992}
\bea\label{eq:psi:transform}
    \psi_0(\chi_0) = \sqrt{|h'(\chi_0)|} \, \psi_Y \left[h(\chi_0) \right] \,,
\eea
where $\psi_Y$ is given in Eq.~(\ref{eq:Psi:Gaussian}). One can check that this transformation both preserves probability $|\psi_Y(Y)|^2 \dd Y = |\psi_0(\chi_0)|^2 \dd \chi_0$ and ensures that $\hat{\pi}_0$ acts as the canonical momentum operator conjugate to $\chi_0$, \ie~$\hat{\pi}_0 \psi_0 = - i \frac{\partial}{\partial \chi_0} \psi_0$. 

Writing out the wavefunction explicitly, one finds
\bea\label{eq:psiCR}
    \psi_0 (\chi_0) = & \left( \frac{2 \kappa^2 \betaR}{\pi} \right)^{1/4} \left( 1 - e^{-2 \frac{|\beta|^2}{\betaR} Y_b^2} \right)^{-1/2} e^{\frac{1}{4} \epsilon_2 H \chi_0} \\
    & \times \left( \exp \left[ - \beta Y_{\rm b}^2 \left( e^{\frac{1}{2} \epsilon_2 H \chi_0} - 1 \right)^2 \right] - \exp \left[ - \beta Y_{\rm b}^2 \left( e^{\frac{1}{2} \epsilon_2 H \chi_0} + 1 \right)^2 \right] \right) \,,
\eea
where the first line comes from the normalisation and Jacobian factors, the second line comes from substituting the non-linear relationship~\eqref{eq:Y:chi0} in the Gaussian wavepackets of~\eqref{eq:Psi:Gaussian}, and we have also used $Y_{\rm b} = - 2 \kappa/\epsilon_2 H$. Due to the non-linear relationship between $Y$ and $\chi_0$, the wavefunction for $\chi_0$ is generically non-Gaussian. In the SR case, one may take the limit $\epsilon_2 \rightarrow 0$, for which $Y_{\mathrm{b}} \rightarrow \infty$, to find
\bea\label{eq:psiSR}
    \psi_0^{\rm SR}(\chi_0) = \left( \frac{2 \kappa^2 \betaR}{\pi} \right)^{1/4} \exp \left( - \kappa^2 \beta \chi_0^2 \right) \,,
\eea
which is simply a complex Gaussian. This occurs because in the SR limit the mapping between the variables $\chi_0$ and $Y$ reduces to linear.

\subsection{Matching conditions}
\label{sec:constant:eps2:matching:conditions}

If $\epsilon_2$ is constant, the general solution of Eq.~(\ref{eq:MS}) is 
\begin{equation}\label{eq:genRsol}
    \mathcal{R}_k(\eta) = \sqrt{\frac{\pi}{2}} \frac{H}{\sqrt{2 \Mp^2 \epsilon_1}} \frac{(- k \eta)^{3/2}}{\sqrt{2 k^3}} \left[ A_k H_\nu^{(1)}(-k \eta) + B_k H_\nu^{(2)}(- k \eta) \right] \,,
\end{equation}
where $H_\nu^{(1)}(x)$ and $H_\nu^{(2)}(x)$ are Hankel functions of the first and second kind, respectively, and $\nu = |3 + \epsilon_2|/2$. The Bogoliubov coefficients $A_k$ and $B_k$ should be normalised as $|A_k|^2 - |B_k|^2 = 1$ in order for $\hat{\mathcal{R}}_{\mathbf{k}}$ and its conjugate momentum to satisfy the canonical commutation relations. Imposing the Bunch-Davies vacuum conditions in the asymptotic past amounts to setting $A_k=1$ and $B_k=0$, but in principle other Bogoliubov coefficients may be considered. Indeed, constant-roll periods with $\epsilon_2 < 0$ cannot extend indefinitely into the past since the first Hubble-flow parameter $\epsilon_1 \propto |\eta|^{-\epsilon_2}$ would grow without bound as $\eta \rightarrow - \infty$, eventually violating the condition for inflation $(\epsilon_1 < 1)$. 

In that case, we imagine that while the modes of interest were still sub-Hubble, there existed an initial slow-roll period which transitioned to constant roll at some $\eta_{\mathrm{t}}$. We demand that this transition occur prior to the matching time $\eta_{t} < \eta_\star$ such that the background is described by constant-roll dynamics for the entirety of the phase described by the homogeneous mode dynamics. While perturbations originate in the Bunch-Davies vacuum during the initial slow-roll phase, the transition to constant roll may excite the state, and from the perspective of the homogeneous mode description, the initial conditions will then look like an excited vacuum state with $B_k\neq 0$. If $\eta_{t} \ll \eta_\star$, \ie~for modes that are deeply sub-Hubble at the time of the transition, one recovers $A_k=1$ and $B_k=0$~\cite{Briaud:2025hra}. Nevertheless, we leave the Bogoliubov coefficients arbitrary at this point for maximal generality.

From Eqs.~(\ref{eq:Pkgendef}) and (\ref{eq:genRsol}), the power spectra are 
\bea\label{eq:PRR:PRdotR}
    \mathcal{P}_{\mathcal{R} \mathcal{R}} (\eta_\star, k) = & \frac{H^2 \sigma^3}{16 \pi \Mp^2 \epsilon_{1\star}} \big| A_k H_\nu^{(1)}(\sigma) + B_k H_\nu^{(2)}(\sigma) \big|^2 \,, \\
    \mathcal{P}_{\mathcal{R} \dot{\mathcal{R}}}(\eta_\star, k) = & \left( \nu - \frac{3+\epsilon_2}{2}\right) H \mathcal{P}_{\mathcal{R} \mathcal{R}} (\eta_\star, k)\\
    & - \frac{H^3 \sigma^4}{16 \pi \Mp^2 \epsilon_{1\star}} \left[ A_k H_\nu^{(1)}(\sigma) + B_k H_\nu^{(2)}(\sigma) \right] \left[ A_k^* H_{\nu-1}^{(2)}(\sigma) + B_k^* H_{\nu-1}^{(1)}(\sigma) \right] \,,
\eea 
where $\epsOnestar = \epsilon_1(\eta_\star)$, and hence from Eq.~(\ref{eq:alphaLpower:spectra})
\begin{equation}\label{eq:alpha:matching:condition}
    \alpha_\star = \frac{i \sigma H}{2} \left[ \frac{A_k^* H_{\nu-1}^{(2)}(\sigma) + B_k^* H_{\nu-1}^{(1)}(\sigma)}{A_k^* H_\nu^{(2)}(\sigma) + B_k^* H_\nu^{(1)}(\sigma)} \right] - \frac{i \nu H}{2} \,.
\end{equation}
In SR and USR, $\nu=3/2$. If we further set $(A_k, B_k) = (1, 0)$, as is the case for $\eta_{t} \ll \eta_\star$, one can show that this leads to
\bea\label{eq:alpha:matching:condition:SR:USR:BD}
    \alpha_\star = \frac{iH}{2} \left( \frac{\sigma^2}{1+i\sigma}-\frac{3}{2} \right) \,.
\eea

To determine the initial condition for $\beta_\star$, we can perform the following conversion. The variables $X$ in Eq.~(\ref{eq:x:from:chi}) and $Y$ in Eq.~(\ref{eq:Ycov}) are related as $X = e^{\frac{1}{2}(3+\epsilon_2) H t} Y$ and since $\pi_X = \dot{X}$ and $\pi_Y = e^{(3+\epsilon_2) H t} \dot{Y}$, their conjugate momenta are related as $\pi_X = e^{- \frac{1}{2} (3 + \epsilon_2) H t} \pi_Y + \frac{1}{2} (3 + \epsilon_2) H e^{\frac{1}{2} (3+ \epsilon_2) H t} Y$. The definition of $\alpha$ in Eq.~(\ref{eq:alpha:moments}) can then equivalently be expressed in terms of moments of $Y$ and $\pi_Y$ as
\bea
    \alpha = - \frac{i}{2} \frac{\langle \hat{Y} \hat{\pi}_Y \rangle}{\langle \hat{Y}^2 \rangle} e^{- (3 + \epsilon_2) H t} - \frac{i}{4} (3+\epsilon_2) H \,.
\eea
Meanwhile, since in the perturbative regime before the matching time $t < t_\star$ the wavefunction of $Y$ is still Gaussian, $\psi_Y = (2\betaR/\pi)^{1/4} \, e^{- \beta Y^2}$, one can express $\beta$ in terms of the moments of $Y$ and $\pi_Y$ analogously to Eq.~(\ref{eq:alpha:moments}), \ie
\bea
    \beta = - \frac{i}{2} \frac{\langle \hat{Y} \hat{\pi}_Y \rangle}{\langle \hat{Y}^2 \rangle} \,.
\eea
Combining with the above expression for $\alpha$, the parameters are related as
\bea\label{eq:beta:alpha:relation}
    \beta = \left[ \alpha + \frac{i}{4} (3+\epsilon_2) H \right] e^{(3+\epsilon_2) H t} \,,
\eea
and in particular at matching, using Eq.~(\ref{eq:alpha:matching:condition:SR:USR:BD}) one has\footnote{Since this is derived using Eq.~(\ref{eq:alpha:matching:condition:SR:USR:BD}), which is only valid for $\nu = 3/2$, we emphasise that this expression for $\beta_\star$ is only applicable for SR $(\epsilon_2 = 0)$ and USR $(\epsilon_2 = -6)$.}
\bea\label{eq:betastar:BD:SR_USR}
    \beta_\star = \frac{i H}{2} \left( \frac{\sigma^2}{1 + i \sigma} + \frac{\epsilon_2}{2} \right) e^{(3 + \epsilon_2) H t_\star} \,.
\eea

\subsection{Wigner function}

We are now in a position to compute the Wigner function, introduced in Eq.~(\ref{eq:Wgendef}). Evaluating this expression for the pure state $\hat{\rho} = \ket{\psi} \! \! \bra{\psi}$, the general definition reads
\begin{equation}
    W(\chi_0,\pi_0) = \frac{1}{2\pi} \int_{-\infty}^{\infty} \dd u \, e^{- i \pi_0 u} \psi^* \left( \chi_0 - \frac{u}{2} \right) \psi \left( \chi_0 + \frac{u}{2} \right) \,.
\end{equation}
For the SR wavefunction of Eq.~(\ref{eq:psiSR}), this evaluates to
\bea\label{eq:WignerSR}
    W_{\rm SR}(\chi_0, \pi_0) = \frac{1}{\pi} \exp \left[ - 2 \kappa^2 \betaR \chi_0^2 - \frac{1}{2 \kappa^2 \betaR} \left( \pi_0 + 2 \kappa^2 \betaI \chi_0 \right)^2 \right] \,.
\eea
We remark that this is both Gaussian and precisely of the form one would expect: In the SR case, the canonical transformation is linear, $Y = \kappa \chi_0$ and $\pi_Y = \pi_0/\kappa$, and so the functional form of the Wigner function is preserved up to coordinate transformations\footnote{The set of linear canonical transformations on phase space $\mathbb{R}^2$ forms the symplectic group $S_p(2,\mathbb{R})$. In quantum mechanics, these correspond to unitary operators on Hilbert space that implement the transformation of states and observables~\cite{Moshinsky:1971rn,Blaszak:2012mqu,Grain:2019vnq}.} -- hence why the Wigner function has the structure $\propto \exp[ - 2 \betaR Y^2 - \frac{1}{2 \betaR} (\pi_Y + 2 \betaI Y)^2]$.

This will not be true of more generic non-linear transformations, as in the case of constant roll with $\epsilon_2 \neq 0$. From the form of the wavefunction in Eq.~(\ref{eq:psiCR}), one finds
\bea\label{eq:WignerCR}
    W(\chi_0,\pi_0) = \frac{1}{2\pi} \left(\! \frac{2 \kappa^2 \betaR}{\pi} \! \right)^{1/2} \! 
    \bigg(\! 1 - e^{- 2 \frac{|\beta|^2}{\betaR} Y_{\mathrm{b}}^2}\! \bigg)^{-1} e^{- 2 \betaR Y_{\rm b}^2} \, e^{\frac{1}{2} \epsilon_2 H \chi_0} \, \mathcal{I}(\chi_0, \pi_0) \,,
\eea
where
\bea
    \mathcal{I} = \int_{- \infty}^{\infty} \! \dd u \, \exp \left[ - i \pi_0 u - Y_{\rm b}^2 e^{\epsilon_2 H \chi_0} \left( \beta e^{\frac{1}{2} \epsilon_2 H u} + \beta^* e^{- \frac{1}{2} \epsilon_2 H u} \right) \right] \big\{ E_1 + E_2 + E_3 + E_4 \big\} \,,
\eea
and 
\bea
    E_1 & = \exp \left[ 2 Y_{\rm b}^2 e^{\frac{1}{2} \epsilon_2 H \chi_0} \left( \beta e^{\frac{1}{4} \epsilon_2 H u} + \beta^* e^{- \frac{1}{4} \epsilon_2 H u} \right) \right] \,, \\
    E_2 & = - \exp \left[ - 2 Y_{\rm b}^2 e^{\frac{1}{2} \epsilon_2 H \chi_0} \left( \beta e^{\frac{1}{4} \epsilon_2 H u} - \beta^* e^{- \frac{1}{4} \epsilon_2 H u} \right) \right] \,, \\
    E_3 & = - \exp \left[ 2 Y_{\rm b}^2 e^{\frac{1}{2} \epsilon_2 H \chi_0} \left( \beta e^{\frac{1}{4} \epsilon_2 H u} - \beta^* e^{- \frac{1}{4} \epsilon_2 H u} \right) \right] \,, \\
    E_4 & = \exp \left[ - 2 Y_{\rm b}^2 e^{\frac{1}{2} \epsilon_2 H \chi_0} \left( \beta e^{\frac{1}{4} \epsilon_2 H u} + \beta^* e^{- \frac{1}{4} \epsilon_2 H u} \right) \right] \,.
\eea
Terms $E_1$ and $E_4$ correspond to the original and reflected Gaussian, respectively, while $E_2$ and $E_3$ represent cross terms, from which one might anticipate interference. 
By performing the change of variables $w = e^{\frac{1}{4} \epsilon_2 H u}$, the integral $\mathcal{I}$ can be brought into the form
\bea\label{eq:Iwvariable}
    \mathcal{I} = \frac{4}{|\epsilon_2| H} \int_0^\infty \dd w \, w^{- \frac{4 i \pi_0}{\epsilon_2 H} - 1} \exp \big[ - Y_{\rm b}^2 e^{\epsilon_2 H \chi_0} \left( \beta w^2 + \beta^* w^{-2} \right) \big] \sum_{i=1}^4 E_i \,,
\eea
with
\bea
    \sum_{i=1}^4 E_i = 4 \sinh \left( 2 Y_{\rm b}^2 e^{\frac{1}{2} \epsilon_2 H \chi_0} \beta w \right) \sinh \left( 2 Y_{\rm b}^2 e^{\frac{1}{2} \epsilon_2 H \chi_0} \beta^* w^{-1} \right) \,.
\eea
This integral does not have an exact closed form solution, and so must be evaluated numerically.\footnote{It is possible to express $\mathcal{I}$ in terms of an infinite series, which will aid in the interpretation of features in Sec.~\ref{sec:results}. We do so in Eq.~\eqref{eq:Iexplicit}.} We do so in the next section, where we study the behaviour of the Wigner function as the state evolves.

\subsection{Comparison with linear theory}\label{sec:linearcompare}

In order to verify the consistency of the above framework, let us check that standard results of cosmological perturbation theory are recovered in the linear limit. At leading order in $\chi_0$, the function $h$ appearing in Eq.~\eqref{eq:chi0_pi0:Y_piY} is linear, namely $h_{\mathrm{lin}}(\chi_0)=\kappa\chi_0$, hence the wavefunction of Eq.~(\ref{eq:psiCR}) reduces to the Gaussian form
\bea\label{eq:psilin}
    \psi_{\mathrm{lin}} (\chi_0) = & \left( \frac{2 \kappa^2 \betaR}{\pi} \right)^{1/4} \exp \left( - \kappa^2 \beta \chi_0^2  \right) \,,
\eea
where we have also used Eq.~\eqref{eq:Y2:Yb:comp} to discard terms that are exponentially suppressed by the inverse squared amplitude of the curvature perturbation. This expression coincides exactly with the SR expression of Eq.~(\ref{eq:psiSR}). It is then not surprising that the Wigner function for the linearised state coincides also with the SR expression~\eqref{eq:WignerSR},
\bea\label{eq:Wignerlin}
    W_{\mathrm{lin}}(\chi_0, \pi_0) = \frac{1}{\pi} \exp \left[ - 2 \kappa^2 \betaR \chi_0^2 - \frac{1}{2 \kappa^2 \betaR} \left( \pi_0 + 2 \kappa^2 \betaI \chi_0 \right)^2 \right] \,.
\eea
One thus obtains a centred Gaussian state, with variances and covariance
\bea
    \langle \hat{\chi}_0^2\rangle = & \frac{1}{4 \kappa^2 \betaR} \, , \quad
    \langle \hat{\pi}_0^2\rangle =   \kappa^2 \frac{\vert\beta\vert^2}{\betaR} \, , \quad
    \langle \hat{\chi}_0 \hat{\pi}_0 \rangle =  \frac{i}{2}-\frac{\betaI}{2\betaR} \,,
\eea 
where $\beta$ is given in Eqs.~\eqref{eq:beta} and~\eqref{eq:betastar:BD:SR_USR}. We remark that the determinant of the covariance matrix is $1/4$, as one would expect for a minimum-uncertainty Gaussian state.
In the late-time limit $t \rightarrow \infty$ and working to leading order in $\sigma \ll 1$, the real and imaginary parts of $\beta$ have the following asymptotic behaviours for SR and USR: 
\bea
    \betaR(t \rightarrow \infty) & = \begin{cases} \frac{1}{2} H \sigma^3 e^{3Ht_\star} & \text{(SR)} \,, \\ \frac{1}{2} H\sigma^3 e^{3H(t_\star-2t)} & \text{(USR)} \,, \end{cases} \\
    \betaI(t \rightarrow \infty) & = \begin{cases} \frac{1}{2} H\sigma^2 e^{3Ht_\star} & \text{(SR)} \,, \\ - \frac{3}{2} H e^{-3Ht} & \text{(USR)} \,. \end{cases} 
\eea 
Together with the definition of $\kappa$ in Eq.~(\ref{eq:Y:chi0}) and $\mathcal{V}$ in Eqs.~(\ref{eq:mu:def}) and~\eqref{eq:mu2:def}, this leads to the following late-time limits for the variances and covariance: 
\bea\label{eq:chi02:PT}
    \langle \hat{\chi}_0^2 \rangle \big|_{t \rightarrow \infty} = \begin{cases} \frac{1}{8 \pi^2 \Mp^2 \epsOnestar} & (\text{SR}) \,, \\ \frac{1}{8 \pi^2 \Mp^2 \epsOnestar} \left(\frac{a}{a_\star}\right)^{6} & (\text{USR}) \,, \end{cases}
\eea 
\bea\label{eq:pi02:PT}
    \langle \hat{\pi}_0^2\rangle \big|_{t \rightarrow \infty} = \begin{cases} \frac{2\pi^2 \Mp^2 \epsOnestar}{\sigma^{2}} & (\text{SR}) \,, \\ \frac{18 \pi^2 \Mp^2 \epsOnestar}{\sigma^{6}} & (\text{USR}) \,, \end{cases} 
\eea 
\bea\label{eq:chi0pi0:PT}
    \text{Re}\left[ \langle \hat{\chi}_0 \hat{\pi}_0\rangle \right] \big|_{t \rightarrow \infty} = \begin{cases} -\frac{1}{2\sigma} & (\text{SR}) \,, \\ \frac{3}{2\sigma^3} \left(\frac{a}{a_\star}\right)^{3} & (\text{USR}) \,. \end{cases}
\eea 

Meanwhile, in perturbation theory at leading order in gradients, the mode equation~\eqref{eq:MS} has two homogeneous solutions
\bea\label{eq:Rhomdef}
    \mathcal{R}_{\mathrm{h},k}  = & C_k + D_k \int^\eta_{\eta_\star}\frac{\dd\tilde{\eta}}{z^2(\tilde{\eta})}
    = C_k - \frac{D_k}{a_\star z_\star^2 (3+\epsilon_2)H} \left[ \left( \frac{\eta}{\eta_\star} \right)^{3+\epsilon_2} -1 \right] ,
\eea
often referred to as the ``growing'' and ``decaying'' modes, respectively. Here, the lower bound of the integral has been set to $\eta_\star$ for convenience, but up to a redefinition of $C_k$ and $D_k$, it can be set otherwise.\footnote{In particular, setting the lower limit to zero and defining $\tilde{C}_k = C_k - D_k \int_0^{\eta^\star} \dd \tilde{\eta} / z^2(\tilde{\eta})$ such that $\mathcal{R}_{\mathrm{h},k} = \tilde{C}_k + D_k \int_0^\eta \dd \tilde{\eta}/z^2(\tilde{\eta})$ recovers the usual intuition that at late times in a SR background, the decaying mode has decayed away such that all that remains is the constant ``growing mode'' solution $\tilde{C}_k$.} In the separate-universe approach, the homogeneous solution $\mathcal{R}_{\mathrm{h},k}$ is employed to describe perturbations on sufficiently large super-Hubble scales, while perturbations on smaller scales are described by the full mode function $\mathcal{R}_k$. To ensure consistency between the descriptions, a matching procedure is employed to set $C_k$ and $D_k$ such that the homogeneous solution~\eqref{eq:Rhomdef} and the full mode function~\eqref{eq:genRsol}, as well as their derivatives, coincide at the matching time (see \eg \cite{Jackson:2023} for more details). This leads to the identification
\bea\label{eq:CkDkdefs} 
    C_k = \mathcal{R}_k(\eta_\star) \,, \quad D_k = z_\star^2 \mathcal{R}^\prime_k(\eta_\star) \,.
\eea 
Assuming Bunch-Davies vacuum conditions $(A_k,B_k) = (1,0)$, the super-horizon limit of the full mode function~\eqref{eq:genRsol} is
\bea\label{eq:Rk:late_time:SR}
    \mathcal{R}_k( \eta) \xrightarrow[]{(- k \eta) \ll 1} - \frac{i 2^\nu \Gamma(\nu) \sigma^{\frac{3}{2} - \nu} H}{\sqrt{8 \pi \Mp^2 \epsOnestar k^3}} \left( \frac{\eta}{\eta_\star} \right)^{\frac{3+ \epsilon_2}{2} - \nu} \left[ 1 + \frac{(- k \eta)^2}{4(\nu - 1)} + \mathcal{O}(-k\eta)^4 \right] \,.
\eea
Recalling that $\nu = |3+\epsilon_2|/2$, notice that the overall multiplicative factor $(\eta/\eta_\star)^{\frac{3+\epsilon_2}{2} - \nu}$ vanishes for $\epsilon_2 > - 3$ but not for $\epsilon_2 < -3$, such that the leading-order time dependence is $\mathcal{R}_k \sim \text{constant} + \mathcal{O}( \eta^2)$ in the former case and $\mathcal{R}_k \sim \eta^{3 + \epsilon_2}$ in the latter. One can then identify the leading-order homogeneous matching coefficients~\eqref{eq:CkDkdefs} as
\bea
    C_k & = - \frac{i 2^\nu \Gamma(\nu) \sigma^{\frac{3}{2} - \nu} H}{\sqrt{8 \pi \Mp^2 \epsOnestar k^3}} \,,\quad 
    D_k  = - z_\star^2 C_k \begin{cases} \frac{k \sigma}{2(\nu - 1)} & \epsilon_2 > - 3  \\ \frac{(3 + \epsilon_2) k}{\sigma} & \epsilon_2 < - 3  \end{cases}\,,
\eea
where we have used that $ k \eta_\star = -\sigma$. From these coefficients, we construct the homogeneous mode function according to Eq.~(\ref{eq:Rhomdef}). In the late-time limit $\eta \rightarrow 0^-$ at leading order in $\sigma$, this homogeneous solution approaches
\bea
    \mathcal{R}_{\mathrm{h}, k} (\eta \rightarrow 0^-) = C_k \begin{cases} 1 & \epsilon_2 > -3  \\ \left( \frac{a}{a_\star} \right)^{-(3+\epsilon_2)} & \epsilon_2 < - 3  \end{cases}\,,
\eea
where we have also used $a = - 1/H\eta$ to simplify. We can also construct the momentum conjugate to $\mathcal{R}_{\mathrm{h},k}$ according to $\pi_{\mathrm{h},k} = z^2 \mathcal{R}_{\mathrm{h},k}'$. Using that $z^2 = 2 \Mp^2 \epsOnestar a_\star^2 (a/a_\star)^{2 + \epsilon_2}$, one can show that in the late-time limit it approaches
\bea
    \pi_{\mathrm{h},k} (\eta \rightarrow 0^-) = - 2 C_k \Mp^2 \epsOnestar a_\star^2 \begin{cases} \frac{k \sigma}{2(\nu-1)} & \epsilon_2 > -3 \,, \\ \frac{(3+\epsilon_2) k}{\sigma} & \epsilon_2 < -3 \,, \end{cases}
\eea
which is constant for both cases. In particular when $\nu = 3/2$ (\ie~for SR or USR), one has
\bea
    |\mathcal{R}_{\mathrm{h},k}^2\big|_{\eta \rightarrow 0^-} = \frac{2\pi^2}{k^3} \begin{cases} \frac{H^2}{8\pi^2 \Mp^2 \epsOnestar} & (\text{SR}) \,,\\
    \frac{H^2}{8\pi^2 \Mp^2 \epsOnestar} \left( \frac{a}{a_\star} \right)^6 & (\text{USR}) \,, \end{cases}
\eea
\bea
    |\pi_{\mathrm{h},k}^2\big|_{\eta \rightarrow 0^-} = \frac{k^3}{2\pi^2} \begin{cases} \frac{2\pi^2 \Mp^2 \epsOnestar}{\sigma^2 H^2} & (\text{SR}) \,,\\
    \frac{18 \pi^2 \Mp^2 \epsOnestar}{\sigma^6 H^2} & (\text{USR}) \,, \end{cases}
\eea
\bea
    \big( \mathcal{R}_{\mathrm{h},k} \pi_{\mathrm{h}, k}^* \big) \big|_{\eta \rightarrow 0^-} = \begin{cases} - \frac{1}{2 \sigma} & (\text{SR}) \,, \\ \frac{3}{2 \sigma^3} \left( \frac{a}{a_\star} \right)^3 & (\text{USR}) \,. \end{cases}
\eea
Comparing with the expressions in Eqs.~\eqref{eq:chi02:PT},~\eqref{eq:pi02:PT}, and~\eqref{eq:chi0pi0:PT}, we see that
\bea
    \mu_1 \mu_2 \frac{k^3}{2\pi^2} |\mathcal{R}_{\mathrm{h},k}^2\big| & = H^2 \langle \hat{\chi}_0^2 \rangle \,, \\
    \frac{\mu_1}{\mu_2} \frac{k^3}{2\pi^2} |\pi_{\mathrm{h},k}^2\big|^2 & = \frac{\langle \hat{\pi}_0^2 \rangle}{H^2} \,, \\
    \mu_1 \frac{k^3}{2\pi^2} \big( \mathcal{R}_{\mathrm{h},k} \pi_{\mathrm{h},k}^* \big) & = \text{Re}\left[ \langle \hat{\chi}_0 \hat{\pi}_0 \rangle \right] \,.
\eea
This precisely mimics the replacement rule of Eq.~(\ref{eq:gh:Deltalnk}), which confirms that our formalism reduces to the standard separate-universe approach in the perturbative limit. In this sense, the strategy proposed in this work may be seen as a quantum version of the separate-universe, or $\delta N$, picture.\\ 

\noindent \textbf{Squeezing}\\  

\noindent Before studying the Wigner function at the non-perturbative level, a word is in order regarding quantum squeezing. In the SR case, from Eqs.~\eqref{eq:chi02:PT}-\eqref{eq:chi0pi0:PT} it is clear that the covariance matrix in the plane $(\chi_0,\pi_0)$ freezes, and hence so does the squeezing parameter
\bea
    r = \frac{1}{2} \cosh^{-1} \! \left( \langle \hat{\chi}_0^2 \rangle + \langle \hat{\pi}_0^2 \rangle \right) \,.
\eea
This appears to contradict the standard lore that quantum squeezing takes place at super-Hubble scales. However, there is no real inconsistency. At leading order in the gradient expansion, and at leading order in cosmological perturbation theory, the action~\eqref{eq:action} does not involve $\chi$ explicitly, hence the momentum is conserved. This is reflected in the fact that the variance $\langle \hat{\pi}_0^2\rangle$ of Eq.~\eqref{eq:pi02:PT} is constant, both in SR and USR. Moreover, in the presence of a dynamical attractor, the curvature perturbation is conserved at super-Hubble scales~\cite{Lyth:2004gb}, which explains why $\langle \hat{\chi}_0^2\rangle$ is conserved in the SR case -- see Eq.~\eqref{eq:chi02:PT}. We thus conclude that in SR, squeezing does not occur in the $(\chi_0,\pi_0)$ phase-space at leading order in gradients.

There are however two caveats that are worth mentioning. First, squeezing depends on the coordinates chosen to parameterise phase-space, and hence using different canonical variables would lead to different conclusions. Squeezing is neither observable nor an intrinsic property of the quantum state for time-dependent Hamiltonians~\cite{Agullo:2022ttg}. Second, the full mode function~\eqref{eq:genRsol} is such that $\mathcal{R}'_k \propto a^{-1}$ at late times in SR, as can be seen from Eq.~\eqref{eq:Rk:late_time:SR}. Hence, the full conjugated momentum $\pi_{\mathcal{R},k} = z^2\mathcal{R}_k' \propto a$ does \textit{not} freeze, unlike the homogeneous version $\mathcal{\pi}_{\mathrm{h},k}$. The reason for this can be traced back to the presence of the first gradient correction to the growing mode in Eq.~\eqref{eq:Rk:late_time:SR} (\ie~the term in parenthesis $\propto (- k \eta)^2$), which happens to overtake the homogeneous decaying mode $\propto (- k \eta)^3$ at late times in SR. Because this gradient correction is not present in the homogeneous description, $\mathcal{R}_{\mathrm{h},k}' \propto a^{-2}$ and thus $\pi_{\mathrm{h},k} = z^2 \mathcal{R}_{\mathrm{h},k}' \propto \text{constant}$ in SR. The absence of squeezing in the $(\chi_0,\pi_0)$ plane thus reflects the fact that the separate-universe approach only keeps track of the homogeneous mode, and that squeezing in these coordinates arise from higher-gradient effects. We further discuss how these gradient corrections could be included in our formalism in Sec.~\ref{sec:conclusions}. 

Finally, note that in the USR case, for which the roles of the ``growing'' and ``decaying'' modes are switched, the mode which grows scales as $\mathcal{R}_k \propto a^3$, and hence $\pi_{\mathcal{R},k} = z^2 \mathcal{R}_k' \propto \text{constant}$ is frozen. Because the gradient correction is always subdominant, the same is true for the homogeneous description: $\mathcal{R}_{\mathrm{h},k} \propto a^3$ and $\pi_{\mathrm{h}, k} \propto \text{constant}$. Consequently, for USR we expect to obtain the same squeezing in the $(\chi_0,\pi_0)$ plane as occurs in standard cosmological perturbation theory. 

\section{Results}\label{sec:results}

\begin{figure}[t!]
\centering
\includegraphics[width=0.6\textwidth]{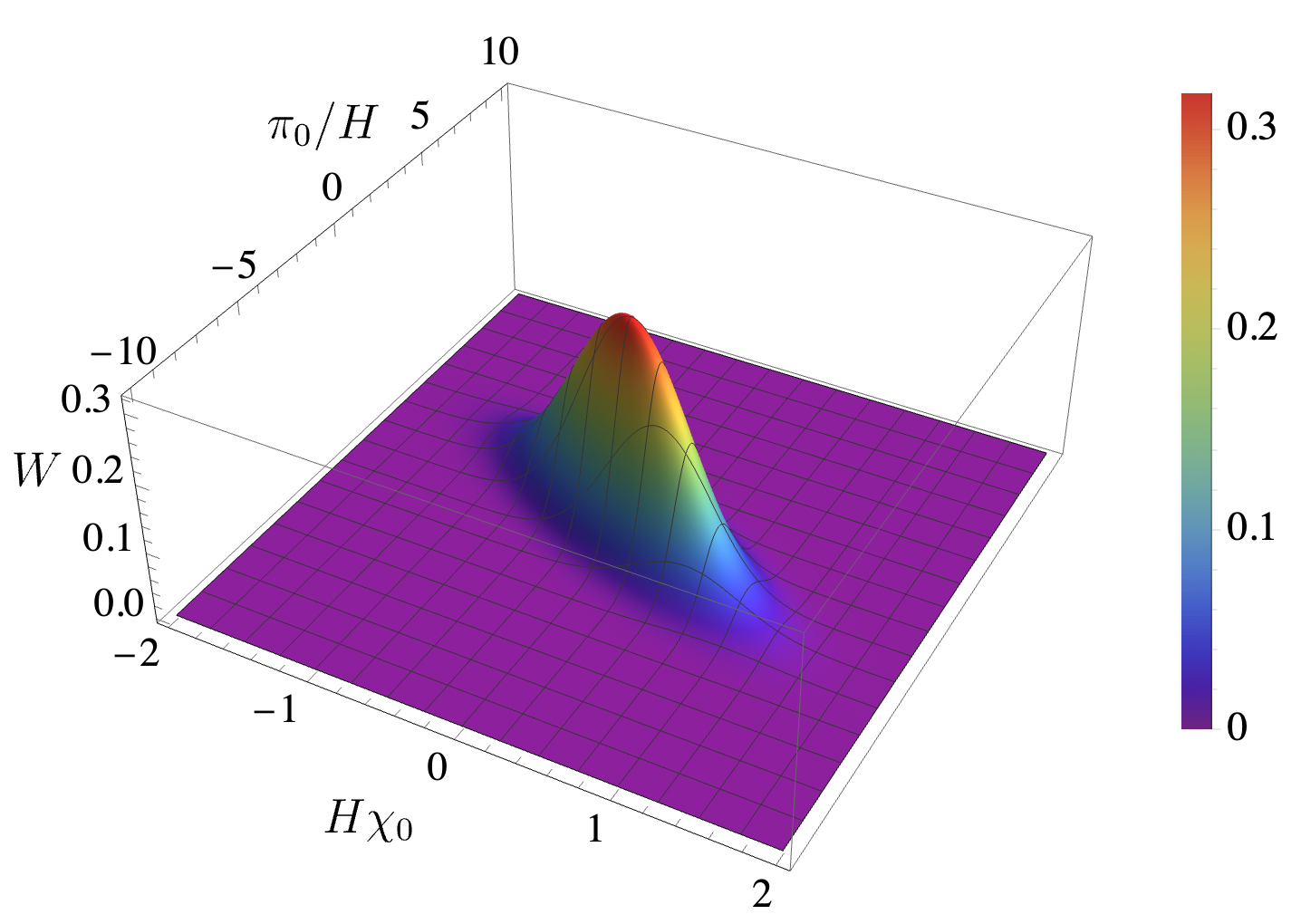}
\caption{Wigner function in a SR background for a representative benchmark -- RB in Eq.~\eqref{eq:RB} -- evaluated at $\Delta N_\star \equiv N - N_\star = 0$ $e$-folds after the matching time. Cross sections correspond to ellipses in phase space centred at the origin. Note that we plot the dimensionless combinations $H \chi_0$ and $\pi_0/H$.}
\label{fig:SRWigner}
\end{figure}
For the sake of comparison, we begin by plotting in Fig.~\ref{fig:SRWigner} the Wigner function in a SR background~\eqref{eq:WignerSR}. As shown in the previous section, $W_{\rm SR}(\chi_0,\pi_0)$ also coincides with the linear-theory result $W_{\rm lin}(\chi_0,\pi_0)$ given in Eq.~\eqref{eq:Wignerlin}. As expected, sections of the Gaussian state trace out ellipses in phase space centred at the origin.
In principle, the ellipticity is controlled by the squeezing parameter $r$ while the orientation is determined by the squeezing angle. As discussed previously, however, there is no squeezing in the homogeneous description of the SR background since it comes from the gradient correction to the growing mode, which is absent in this description. Consequently, there is little to be gained from analysing dynamics in the homogeneous description of SR further. 

\begin{figure}[t!]
\centering
\includegraphics[width=1.0\textwidth]{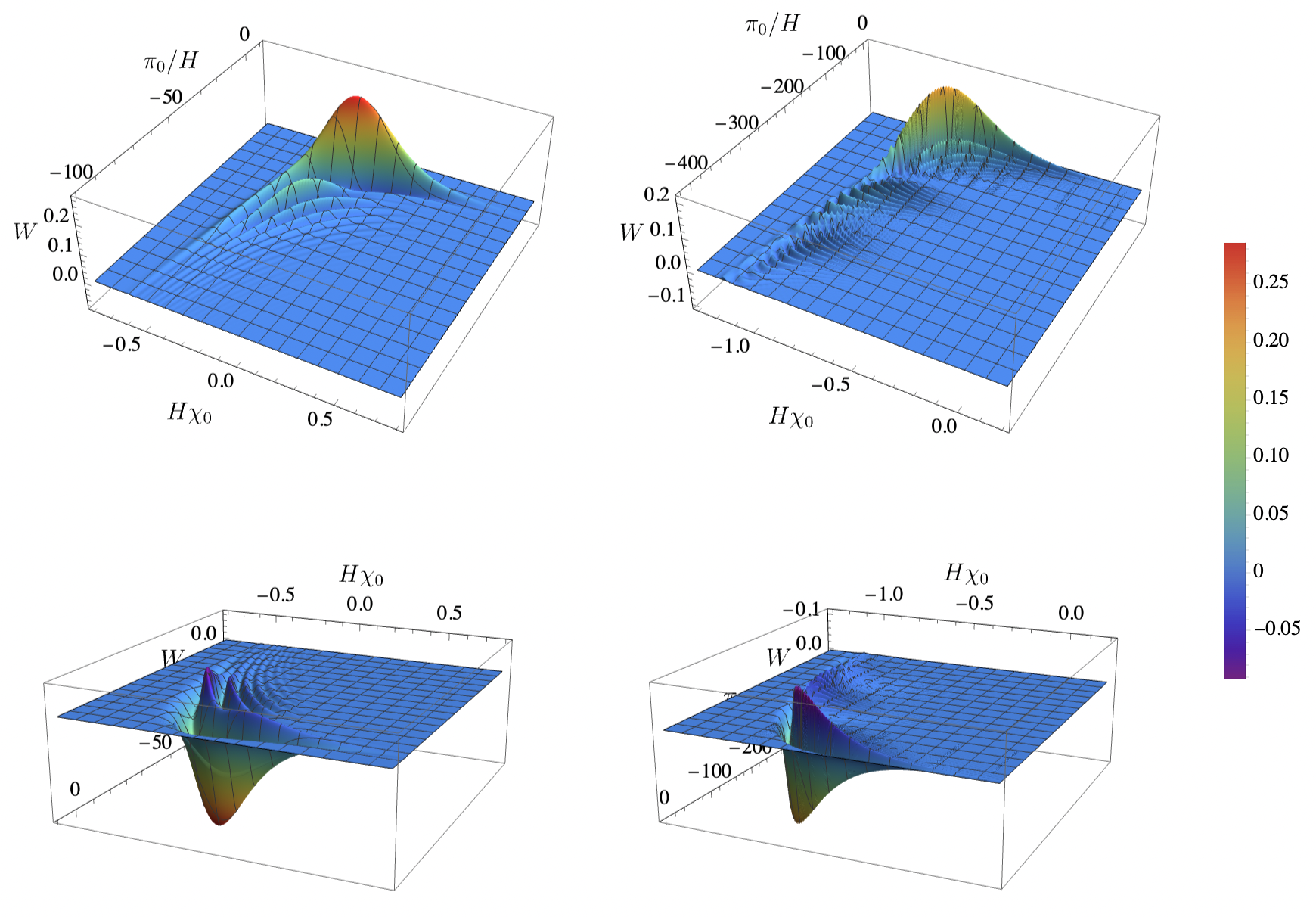}
\caption{Wigner function in a USR ($\epsilon_2 = -6$) background for the reference benchmark in Eq.~\eqref{eq:RB} evaluated at $\Delta N_\star = 0$ (left column) and $\Delta N_\star = 0.72$ (right column) $e$-folds after the matching time. The bottom panels show a different angle from which it is clear that the Wigner function assumes non-positive values.}
\label{fig:USRWigner}
\end{figure}

\subsection{Wigner oscillations and interference fringes}

We now turn to the more general constant-roll Wigner function, focusing in particular on the USR case ($\epsilon_2 = -6$). In Fig.~\ref{fig:USRWigner}, we numerically evaluate Eq.~(\ref{eq:WignerCR}) for our main reference benchmark (RB) with\footnote{In Appendix~\ref{app:sigma}, we demonstrate that our results are robust to modest variations in $\sigma$.
The relatively large value taken here is chosen for numerical convenience.}
\bea\label{eq:RB}
    \textbf{RB:} \,\,\, (\bar{\mathcal{P}}_{\mathcal{R}},\sigma) = (0.14,\, 1) \,,
\eea
at $\Delta N_\star = 0$ (left panels) and $\Delta N_\star = 0.72$ (right panels), where $\Delta N_\star \equiv N - N_\star$ denotes the number of $e$-folds after the matching time. A benchmark is completely specified by just two parameters: the amplitude of curvature perturbations at the Hubble-crossing time $t=0$, $\bar{\mathcal{P}}_{\mathcal{R}} \equiv \mathcal{P}_{\mathcal{R}\mathcal{R}}(t=0,k) = H^2/(8\pi^2 \Mp^2 \bar{\epsilon}_1)$, and the coarse graining parameter $\sigma$, which sets the matching time via $N_\star = - \ln(\sigma)$. Indeed, from the expressions derived in Sec.~\ref{sec:constant roll}, one can readily check that when $\chi_0$ and $\pi_0$ are rescaled with $H$, only the two parameters above appear in the Wigner function.\footnote{That is, in analogy with the $e$-folding number $N = H t$, we rescale the Goldstone of broken time translations as $\chi_0 \rightarrow H \chi_0$ and its conjugate momentum as $\pi_0 \rightarrow \pi_0/H$.}
The parameters are chosen such that the perturbativity condition is marginally satisfied at the matching time, 
hence the Wigner function already exhibits clear departures from Gaussianity at that time.\footnote{When decreasing the value of $\bar{\mathcal{P}}_{\mathcal{R}}$, the Wigner function looks more Gaussian initially, but it then becomes numerically more difficult to track the appearance of non-Gaussian features.}
Although the distribution remains relatively localised about a central peak near the origin, its previously elliptical profile has been noticeably deformed into a boomerang-shaped contour, with interference fringes extending into one quadrant. Crucially, these oscillations locally dip into negative values, so the Wigner function is no longer globally positive. The same benchmark evaluated at a later time (right panels) reveals the emergence of a cascade of rapidly oscillating interference fringes at very negative values of $\pi_0$, far out in phase space. The envelope of these fringes appears as a repeated, mirrored replica of the central structure, while the ripples themselves exhibit decreasing amplitude and increasing oscillation frequency.

There are in principle two sources contributing to these oscillatory features: 1) interference between the original and reflected Gaussians in~\eqref{eq:Psi:Gaussian}, and 2) the non-linear mapping between $(Y,\pi_Y)$ and the phase-space coordinates of interest $(\chi_0,\pi_0)$.

\subsection*{Interference with the reflected Gaussian}

Though each Gaussian alone would produce a strictly positive Wigner function in the $(Y,\pi_Y)$ phase space, the cross terms give rise to interference in the $Y$-direction (and hence the $\chi_0$-direction after transforming back to the original variables). The nature of this interference is somewhat similar to that seen in Schr{\"o}dinger cat states in quantum optics~\cite{Dodonov:1974mcv}, which are formed from the superposition of two opposite-phase (or opposite-displacement) coherent states $\ket{\psi_{\rm cat}} \propto \left( \ket{\alpha} + e^{i \theta} \ket{- \alpha} \right)$. The superposition of ``position-space'' Gaussians in~\eqref{eq:Psi:Gaussian} is morally similar, and upon computing\footnote{We emphasise that one must compute $W(Y,\pi_Y)$ directly from $\psi_Y(Y)$, as the non-linear change of variables cannot be directly performed at the level of the Wigner function~\eqref{eq:WignerCR}.} the Wigner function corresponding to these phase-space variables $W(Y,\pi_Y)$, one observes a similar interference pattern, with an alternating series of positive and negative lobes. There is a crucial difference, however -- because coherent states are defined on the full line, the interference fringes for the cat state are symmetric and regular. By contrast, because $Y$ is defined on a truncated domain, the result is highly asymmetric. Further, the origin of the second term in the superposition is fundamentally different. The cat state is prepared via an external operation (one uses the annihilation operator to remove one or more photons from a squeezed vacuum state~\cite{Takase:2020gpy}) and so the two terms correspond to physical states. Whereas in our case, the reflected Gaussian feature arises to enforce the boundary condition at $Y_{\rm b}$, namely the vanishing of the wavefunction at $\chi_0\to\pm\infty$. In this sense, it acts analogously to an ``image wavefunction'' in quantum mechanics, which arises from reflection at the boundary and serves to ensure destructive interference there, rather than representing an independently prepared component of a state. 
After performing the change of variables back to $(\chi_0,\pi_0)$, the asymmetric positive and negative lobes become additional wavelet structure on top of the envelope seen in Fig.~\ref{fig:USRWigner}. Let us finally note that the boundary condition is imposed in the asymptotic past and future, $t+\chi_0\to\pm\infty$, which in principle requires to embed our setup in a realistic cosmological scenario. For instance, while we have considered an infinite USR phase, it is clear that it cannot extend indefinitely in the future since inflation must end at some point. How this modifies boundary conditions remains to be clarified, although we expect reflected Gaussian features to be present in general.

\subsection*{Non-linear mapping}

While the first source of interference is present already at the level of the $(Y,\pi_Y)$ variables, there is a second contribution arising from the non-linear canonical transformation~\eqref{eq:chi0_pi0:Y_piY} required to return to the original phase-space variables $(\chi_0,\pi_0)$. That such a transformation should introduce negativity in the Wigner function is closely connected with Hudson's theorem, which says that the only pure quantum state with a globally positive Wigner function is the Gaussian state~\cite{HUDSON1974249}. It is thus inevitable that once the state has been re-cast in the $(\chi_0,\pi_0)$ variables, which renders its non-Gaussian nature manifest, local regions of negativity must appear.\footnote{Wigner negativity is not invariant under non-linear canonical transformations, and thus measures the importance of phase coherence when measurements are performed on particular phase-space variables. Here we focus on $(\chi_0,\pi_0)$ given the fundamental role they play in the EFT approach and their close connection with cosmological observables.} Unlike the contribution described above, this effect is present even in the perturbative regime, where the reflected Gaussian is irrelevant. In this sense, Wigner negativity in constant-roll backgrounds should be viewed as arising not only from non-perturbative boundary effects, but also from features present already at the perturbative level -- non-perturbative methods being, however, required to monitor these features across the full phase-space region where Wigner negativity is realised, as explained in Sec.~\ref{sec:intro}. In phase space, this non-linearity manifests by distorting the simple Gaussian dependence on $\pi_Y$ into an infinite tower of higher harmonics in $\pi_0$, weighted by Bessel functions whose amplitudes decay as one moves further away from the centre of the distribution.\\

To make this last comment more concrete, it is instructive to re-write Eq.~\eqref{eq:Iwvariable} as follows. Using the identities~\cite{NIST:DLMF}
\bea
    e^{\frac{z}{2}( t + t^{-1})} = \sum_{n = -\infty}^\infty t^n I_n(z) \,, \quad e^{\frac{z}{2}( t - t^{-1})} = \sum_{n = -\infty}^\infty t^n J_n(z) \,,
\eea
where $J_n(x)$  and $I_n(x)$ are Bessel and modified  Bessel functions of the first kind respectively, one can rewrite the sum over the $E_i$ terms as the infinite series
\bea
    \sum_{i = 1}^4 E_i = 2 \sum_{n = -\infty}^\infty \left( \frac{\beta}{|\beta|} \right)^{2n} \left[ I_{2n} \left(4 Y_{\rm b}^2 |\beta| e^{\frac{1}{2} \epsilon_2 H \chi_0} \right) - J_{2n}\left(4 Y_{\rm b}^2 |\beta| e^{\frac{1}{2} \epsilon_2 H \chi_0} \right) \right] w^{2n} \,,
\eea
where we have used that $I_n(-z)=(-1)^n I_n(z)$ and $J_n(-z)=(-1)^n J_n(z)$. The integral appearing in~\eqref{eq:Iwvariable} then involves terms of the form
\bea
    \mathcal{I}_n = \int_0^\infty \frac{\dd w}{w} \, w^{- \frac{4 i \pi_0}{\epsilon_2 H} + 2n} \exp[ - Y_{\rm b}^2 e^{\epsilon_2 H \chi_0} (\beta w^2 + \beta^* w^{-2})] \,,
\eea
which can be evaluated with the aid of the identity~\cite{NIST:DLMF}
\bea
    \int_0^\infty \frac{\dd w}{w} \, w^{\nu} e^{-a w - b w^{-1}} = 2 \left( \frac{b}{a} \right)^{\nu/2} K_\nu \left(2 \sqrt{ab} \right) \,,
\eea
where $K_\nu(x)$ is a modified Bessel function of the second kind. Thus, one can recast $\mathcal{I}$ as
\bea\label{eq:Iexplicit}
    \mathcal{I}(\chi_0,\pi_0) = \frac{8}{|\epsilon_2| H} \sum_{n = -\infty}^\infty \bigg[ I_{2n} \bigg(\!4 Y_{\rm b}^2 |\beta| & e^{\frac{1}{2} \epsilon_2 H \chi_0} \!\bigg) - J_{2n}\bigg(\!4 Y_{\rm b}^2 |\beta| e^{\frac{1}{2} \epsilon_2 H \chi_0} \!\bigg) \bigg] \\
    & \times \left( \frac{\beta}{|\beta|} \right)^{\frac{2 i \pi_0}{\epsilon_2 H}+n} K_{- \frac{2 i \pi_0}{\epsilon_2 H} + n}\left(2 Y_{\rm b}^2 |\beta| e^{\epsilon_2 H \chi_0} \right) \,,
\eea
which enters into $W$ in Eq.~\eqref{eq:WignerCR}. In particular, defining for conciseness $x = - \epsilon_2 H \chi_0/2$, $p = - 2 \pi_0/\epsilon_2 H$, $\mathcal{A} = 2 Y_{\rm b}^2 |\beta|$, and $\beta = |\beta| e^{i \phi_\beta}$, we have
\bea\label{eq:Wxp}
    W(x,p) = \mathcal{M} e^{-x} \sum_{n=-\infty}^\infty \left[ I_{2n}(2 \mathcal{A} e^{-x}) - J_{2n}(2 \mathcal{A} e^{-x})\right] e^{i \phi_\beta (n - i p)} K_{n + i p} (\mathcal{A} e^{-2x}) \,,
\eea
with 
\bea
\label{eq:calN:W:def}
    \mathcal{M} = \frac{1}{2\pi} \frac{8}{|\epsilon_2|} \left(\! \frac{2 \kappa^2 \betaR}{\pi} \! \right)^{1/2} \!  \bigg(\! 1 - e^{- 2 \frac{|\beta|^2}{\betaR} Y_{\mathrm{b}}^2}\! \bigg)^{-1} e^{- 2 \betaR Y_{\rm b}^2} \,.
\eea

Writing the Wigner function in this form offers not just a practical advantage, since in practice only a finite number\footnote{When truncating the sum at finite $n$, care must be taken to include both the positive and negative terms in each $\pm n$ pair so that the result is real, as required for the Wigner function.} of terms about $n=0$ need be kept in order to faithfully capture the result of a full numerical evaluation, but it also offers conceptual clarity. Since $I_{2n}(z) - J_{2n}(z) \geq 0$ for real, non-negative argument $z \in \mathbb{R}_{\geq 0}$ and $n \in \mathbb{Z}$, this term cannot generate negativity. The exponential factor\footnote{This $e^{\phi_\beta p}$ factor is the origin of the asymmetry between positive and negative values of $\pi_0 \propto p$. Since $\phi_\beta < 0$ is negative, as can be seen from~\eqref{eq:beta}, large positive values of $\pi_0 \propto p$ are exponentially suppressed, whereas negative values are not.} $e^{i \phi_\beta(n - i p)} = e^{i n \phi_\beta} e^{\phi_\beta p}$ contains a real, strictly positive dependence on the phase-space variable $p$ through $e^{\phi_\beta p}$, and therefore cannot induce oscillations or negativity. The complex phase $e^{i n \phi_\beta}$ \textit{can} generate interference through the sum over $n$, but since it is independent of the phase-space variables, it does not induce oscillatory or sign-changing dependence on $x$ or $p$. Dynamically, oscillations and negativity can arise \textit{only} through $K_{n + i p}(z)$. 
The modified Bessel function of the second kind with complex index $\nu$ is generically oscillatory provided $\text{Im}[\nu] \neq 0$, and the zeroes have been studied in e.g.~\cite{Ferreira:2008}, which provides analytic formulae in the limits $|\nu| \ll 1$ and $|\nu| \gg 1$. Unfortunately these are not particularly useful here, however; while the largest contribution to the infinite sum typically comes from $n=0$, the first few non-zero terms $n=\pm 1, 2, 3 \cdots$ can also contribute appreciably. Thus, the locations of the zeroes are the result of a complex interplay between the different contributions of $K_{n + i p}(z)$ and their weighting factors. 

\begin{figure}[t!]
\centering
\includegraphics[width=0.85\textwidth]{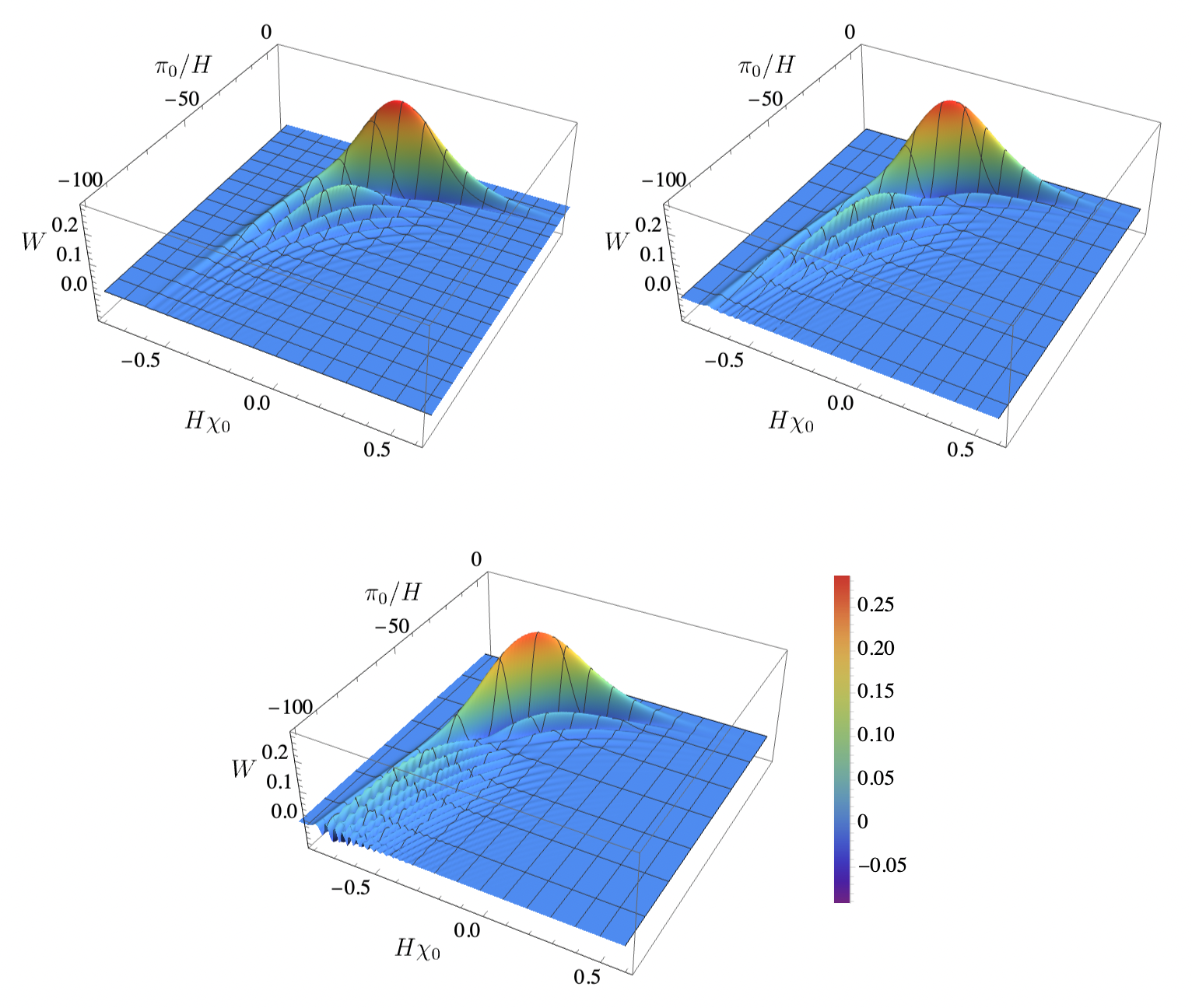}
\caption{USR Wigner function evaluated at three sample time steps: $\Delta N_\star = 0$ (top left), $\Delta N_\star = 0.17$ (top right), and $\Delta N_\star = 0.35$ (bottom), where $\Delta N_\star = N - N_\star$ is the number of $e$-folds after the matching time. Note that at later times, the full distribution extends beyond the small inset shown here (see Fig.~\ref{fig:USRWigner}). Other parameters fixed to RB values -- see Eq.~\eqref{eq:RB}. A video of the time evolution is available \href{https://www.youtube.com/watch?v=sd88gwogfAg}{via this link}.}
\label{fig:USRtimes}
\end{figure}

Nevertheless, a few general statements can be made. In the limit $|\nu| \gg 1 \gg |z|$, the zeroes of $K_\nu(z)$ are roughly exponentially spaced, $z_\ell \sim \nu \exp[- i \pi (\ell - 1/4)/\nu]$, and clustered about $z \sim 0$~\cite{Ferreira:2008}. Since in our case $z \sim e^{- 2 x}$, at large $|p|$ they should appear roughly linear in $x$ and be clustered about small values $|x| \lesssim 1$, which is indeed seen in Fig.~\ref{fig:USRWigner}. Additionally, at fixed $n$, the amplitude of oscillations becomes smaller and the frequency becomes larger, respectively, for larger values of $|\text{Im}[\nu]| = |p|$. This too is reflected in Fig.~\ref{fig:USRWigner}, as the interference fringes become smaller in amplitude but larger in frequency as one moves outwards in phase space to larger values of $|p|$. Finally, we remark that in order for the representation~\eqref{eq:Wxp} with a truncated sum to faithfully capture the behaviour of $W$ at larger values of $|\pi_0| \propto |p|$, one must retain an increasing number of terms in the sum over $n$. In particular, while the low-$n$ terms are sufficient to resolve the smooth pseudo-Gaussian shape near the origin, higher-$n$ terms are required in order to resolve the high-frequency oscillations at large $|\pi_0|$. This is consistent with the interpretation that the non-linear canonical transformation maps a simple Gaussian dependence in the variables $(Y,\pi_Y)$ into an infinite tower of harmonics in $\pi_0$, with higher harmonics progressively activated only at larger momenta.

The form of the Wigner function in Eq.~\eqref{eq:Wxp} also allows us to make sense of the evolution of $W$ with increasing time, as shown in Figs.~\ref{fig:USRWigner} and~\ref{fig:USRtimes}. As time progresses, the distribution becomes less sharply localised about the origin and stretches in the $\pi_0$ direction, reflecting the squeezing discussed at the end of Sec.~\ref{sec:linearcompare}. At later times, additional ripples appear at large, negative values of $\pi_0$. This behaviour can be understood from Eqs.~\eqref{eq:beta} and~\eqref{eq:Wxp} -- with time, the argument of the modified Bessel function $\propto |\beta(t)|$ decreases. In this regime, $K_\nu(z \ll 1)$ features higher-frequency, smaller-amplitude oscillations. The new zeroes correspond to higher values of $n$ in the sum, which are most strongly supported when $|p| \sim |n|$. Additionally, $\phi_\beta$ becomes more negative with advancing time; the factor $e^{\phi_\beta p}$ then suppresses positive values and enhances negative values of the conjugate momentum, effectively pushing the oscillatory structure outwards in $\pi_0 \propto p$.

\subsection{Wigner negativity}

\begin{figure}[t!]
\centering
\includegraphics[width=0.6\textwidth]{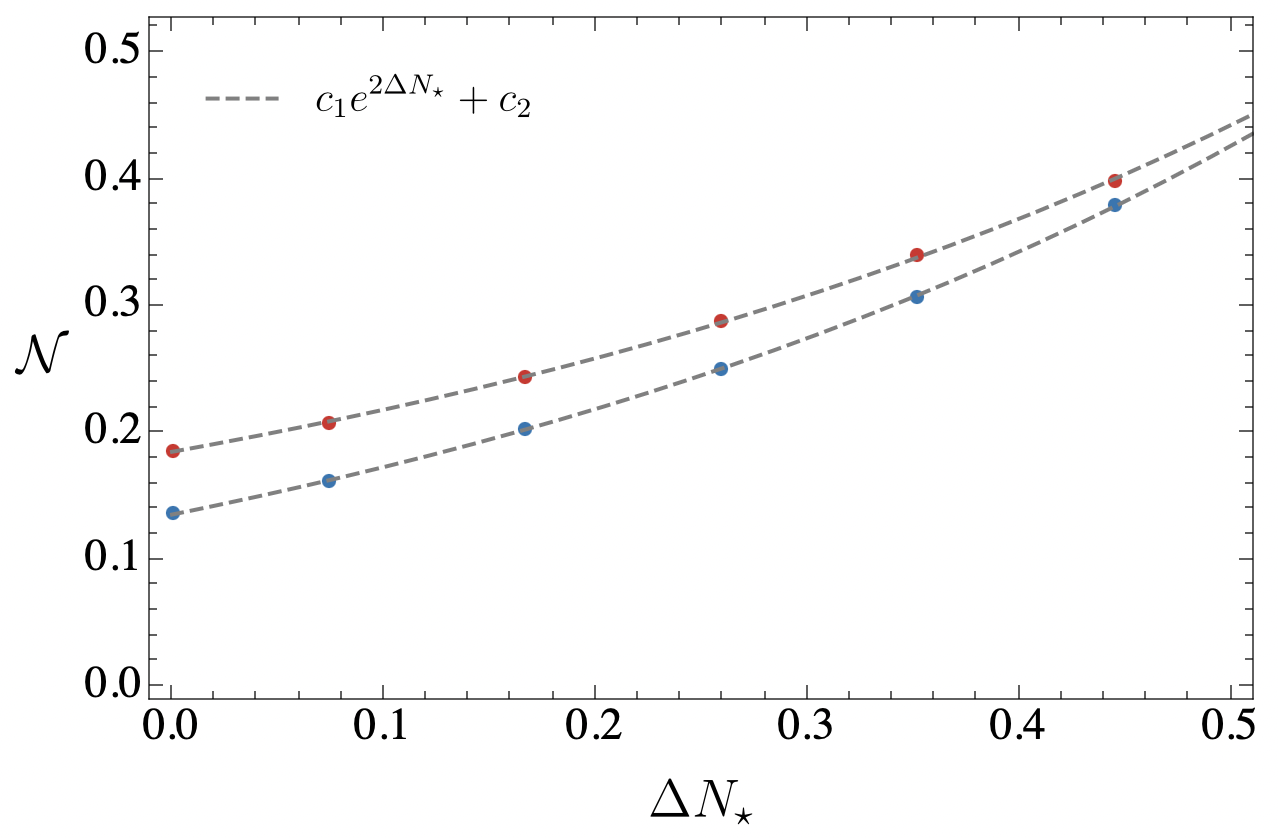}
\caption{Wigner negativity $\mathcal{N}$ of Eq.~\eqref{eq:negativitydef} as a function of the number of $e$-folds since matching, $\Delta N_\star = N - N_\star$, for our reference benchmark of Eq.~\eqref{eq:RB} (red) and a second benchmark with a larger scalar amplitude $(\bar{\mathcal{P}}_\mathcal{R}, \sigma) = (0.28,1)$ (blue). The negativity evolution is well-fit by an exponential growth law $\mathcal{N} = c_1 e^{2 \Delta N_\star} + c_2$. For the reference benchmark, the fitting coefficients are empirically found to be $(c_1, c_2) = (0.15,0.035)$; for the second benchmark, $(c_1, c_2) = (0.17,-0.034)$.}
\label{fig:negativity}
\end{figure}

In order to quantitatively study how the negativity of $W(\chi_0,\pi_0)$ evolves with time, we evaluate the Wigner negativity volume introduced in Eq.~\eqref{eq:negativitydef}. This quantity vanishes for states with positive Wigner functions, while more generally $\mathcal{N} > 0$. The magnitude of $\mathcal{N}$ quantifies the quantum interference encoded in $W(\chi_0,\pi_0)$, with larger values corresponding to more dramatically non-classical states. Conversely, a suppressed $\mathcal{N}$ signals the emergence of an effectively classical phase-space distribution. In Fig.~\ref{fig:negativity} we plot the negativity volume as a function of time, parameterized by the number of $e$-foldings $N = H t$ since the matching time $N_\star$, \ie~$\Delta N_\star \equiv N - N_\star$, for our representative benchmark RB in~\eqref{eq:RB} as well as for a second benchmark with a slightly larger curvature perturbation amplitude at Hubble crossing. 

The first thing to note is that $\mathcal{N}$ grows monotonically with time. The growth law is in principle inherited from the time evolution of $\beta(t)$; however, because $\beta$ enters the Wigner function $W$ in a highly non-linear manner, and because $\mathcal{N}$ itself depends non-linearly on $W$, an analytic derivation of this relationship is not feasible. Instead, fitting the numerically evaluated $\mathcal{N}(\Delta N_\star)$ reveals that, at early times, the growth is well-described by the exponential form
\bea\label{eq:negativityfit}
    \mathcal{N}(\Delta N_\star) = c_1 e^{2 \Delta N_\star} + c_2 \,,
\eea
where the fitting coefficients $(c_1, c_2)$ depend on the benchmark under consideration and must be determined empirically. It is rather suggestive that the negativity initially appears to grow as $e^{2 \Delta N_\star} \propto a^2$. Whether this behaviour persists to later times remains unclear, as numerical methods become difficult to implement efficiently once the function becomes highly oscillatory and squeezing necessitates a denser grid. Nevertheless, the observed growth is clearly convex upwards and shows no indication of saturation over the range readily accessible to our analysis. 

One may be surprised that, in Fig.~\ref{fig:negativity}, larger values for $\bar{\mathcal{P}}_{\mathcal{R}}$ result in smaller negativity: since $\bar{\mathcal{P}}_{\mathcal{R}}$ sets the amplitude of curvature perturbations at Hubble crossing, it also governs the magnitude of non-perturbative effects and, one might therefore expect, the negativity itself. There are however two reasons that explain this seemingly counter-intuitive behaviour. First, extrapolating the empirical growth law in Eq.~(\ref{eq:negativityfit}) to late times reveals that the negativity associated with larger $\bar{\mathcal{P}}_{\mathcal{R}}$ eventually overtakes that of our reference benchmark. Second, it is important to emphasise the there exists both a perturbative region of phase space ($\vert \epsilon_2 H\chi_0\vert <1$) and a non-perturbative one. As $\bar{\mathcal{P}}_{\mathcal{R}}$ decreases, the support of the Wigner function becomes increasingly concentrated in the perturbative region, although it continues to extend over both regions. We have explicitly verified that, when the negativity is computed by restricting to the perturbative region only, smaller values of $\bar{\mathcal{P}}_{\mathcal{R}}$ indeed lead to smaller negativity, in agreement with expectations. See Fig.~\ref{fig:perturbative} in Appendix~\ref{app:perturbative}.

\subsection{Parametric exploration}

\begin{figure}[t!]
\centering
\includegraphics[width=0.9\textwidth]{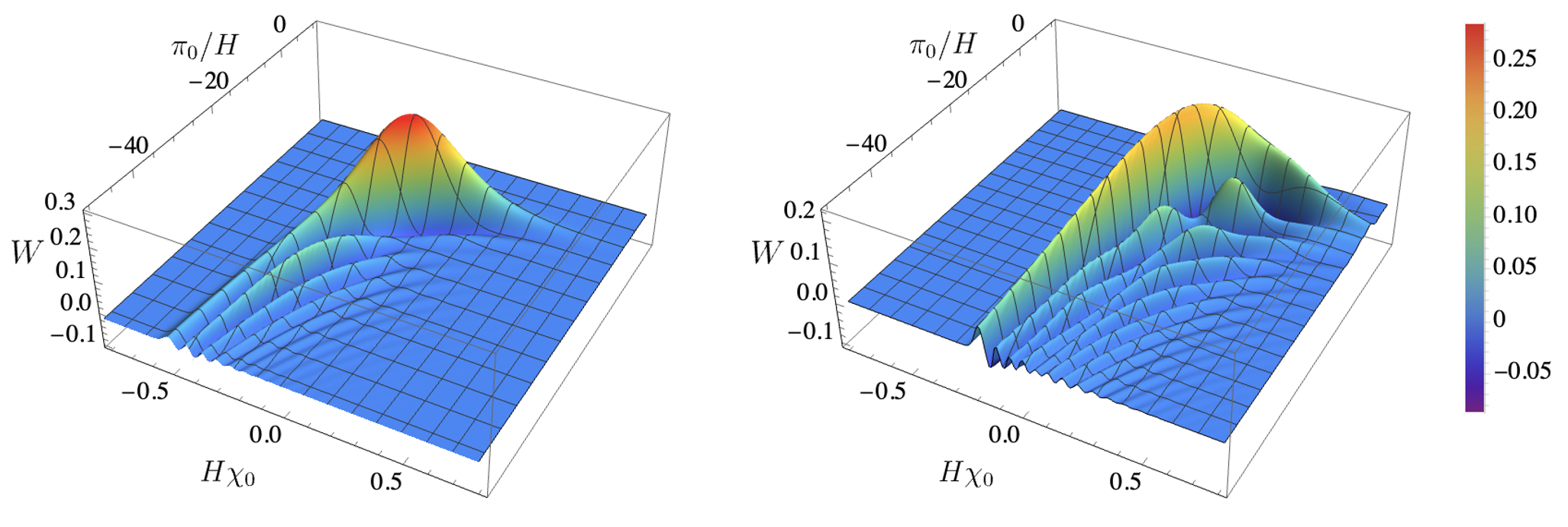}
\caption{\textbf{Left:} USR Wigner function for RB values, $(\bar{\mathcal{P}}_{\mathcal{R}}, \sigma) = (0.14,1)$, at $\Delta N_\star = 0$. \textbf{Right:} Same plot for $(\bar{\mathcal{P}}_{\mathcal{R}}, \sigma) = (0.014,1)$, also evaluated at $\Delta N_\star = 0$.}
\label{fig:USRHeps}
\end{figure}

One can also study how the Wigner function changes upon varying the other parameters thus far held fixed. Fig.~\ref{fig:USRHeps} shows the USR Wigner function for two sample values of $\bar{\mathcal{P}}_{\mathcal{R}}$.
The left panel corresponds to the RB value~\eqref{eq:RB}, while the right panel shows a value smaller by a factor of $10$. For the smaller value, the central boomerang-shaped contour becomes more symmetrically rounded, with additional wavelet structure on top of the envelope seen in the left panel. This behaviour can be understood from the form of Eq.~(\ref{eq:Wxp}) and the fact that $\Mp^2 \epsOnestar/H^2 \propto \mathcal{A}$ appears in the argument of the Bessel functions $I_{2n}(2 \mathcal{A} e^{-x})$, $J_{2n}(2 \mathcal{A} e^{-x})$, and $K_{n+ip}(\mathcal{A} e^{-2 x})$. Increasing $\mathcal{A}$ modifies the balance between these functions, enhancing interference among nearby modes in the series to give rise to the nested wavelet modulations on top of the existing envelope. Further, this larger $\mathcal{A}$ forces the oscillations, which appear for small argument of the modified Bessel $K_{n+ip}(\mathcal{A} e^{-2 x})$, to cluster more closely about $x \propto \chi_0 \sim 0$, as can be seen in the right panel.

Finally, one can also vary the coarse-graining parameter $\sigma$, which controls the scale at which modes are transferred from the short-wavelength sector to long-wavelength effective theory. In our convention, $N_\star = - \ln(\sigma)$, and so decreasing $\sigma$ corresponds to performing the matching at a later time. In the separate-universe approach, physical observables are expected to be independent of the choice of $\sigma$, provided it is taken to be sufficiently small. This is true at the perturbative level; in the present framework, mild $\sigma$-dependence is expected since non-linearities are only included after the matching time. Consequently, taking $\sigma$ to be too small decreases the interval for which non-linear effects are incorporated and therefore underestimates their impact. We have verified that our results are otherwise robust to moderate variations in $\sigma$ (see Fig.~\ref{fig:USRsigma} in Appendix~\ref{app:sigma}).

\section{Discussion}\label{sec:conclusions}

These results have a number of implications. Perhaps the most glaring is that quantum effects \textit{can} become important, potentially even preventing a stochastic description, for inflationary backgrounds that deviate from slow-roll evolution and develop large fluctuations. Recall that a necessary, though not sufficient, condition for formulating a stochastic description in terms of the Wigner function is that this object be everywhere positive, such that its probabilities are meaningful. This condition fails in certain regions of parameter space, as can be seen in the figures of Sec.~\ref{sec:results}. The Wigner negativity can be quite sizeable and generically increases with time (see Fig.~\ref{fig:negativity}). Consequently, $W$ cannot be interpreted as a true probability distribution. It is worth remarking that since $W$ is still positive over certain regions of parameter space, one may still be able to use a stochastic description to reproduce with good accuracy some observables provided their support in phase space coincides with these positive regions. This is probably the case if the amplitude of curvature perturbations remains small and one is interested in measuring low-order correlation functions. However, that the negative regions appear already quite close to the origin in phase space for the USR models studied here bodes well for identifying an observable involving large-density objects (such as ultra-compact dark matter halos or primordial black holes) sensitive to this Wigner negativity, which we leave to future work. 

This finding runs contrary to the standard lore, which holds that the squeezing that occurs while modes are outside the Hubble radius is responsible for their ``classicalisation''. Instead, we find that the squeezing picture -- or equivalently, the hierarchy of linear-theory growing and decaying modes -- is insufficient to diagnose classicality. In retrospect, this should not be too surprising. As we have already remarked, the success of a stochastic description in reproducing the most relevant observables (e.g. 2-point functions of Hermitian observables) is actually attributable to working in the linear theory~\cite{Martin:2015qta}, which preserves a Gaussian state. A squeezed state is still a very quantum state, as is appreciated in quantum optics and other fields; it just appears classical for a certain class of coarse-grained observables. Further, because the squeezing amplitude can always be changed (or even eliminated) by an appropriate canonical transformation, there is no reason why it should be a meaningful reflection of the intrinsic properties of the state~\cite{Agullo:2022ttg}. 

Another important lesson to take away is that slow-roll and ultra-slow-roll backgrounds have surprising qualitative differences. Again, if one follows the schematic squeezing argument,
one would reason that slow roll and ultra-slow roll should be equally classical, since even though what is called the growing and decaying mode switches between the cases, in both backgrounds one mode dominates over the other. We see that this is not so. One may have also expected a certain symmetry between the slow-roll and ultra-slow-roll results on the basis of the Wands duality~\cite{Wands:1998yp}, which relates the spectra of perturbations in cosmological backgrounds with the same Mukhanov-Sasaki mass term $z''/z$, which can be expressed exactly in terms of Hubble-flow parameters as
\bea
    \frac{z''}{z} = \mathcal{H}^2 \left( 2 - \epsilon_1 + \frac{3}{2} \epsilon_2 - \frac{1}{2} \epsilon_1 \epsilon_2 + \frac{1}{4} \epsilon_2^2 + \frac{1}{2} \epsilon_2 \epsilon_3 \right) \,.
\eea
Because for both slow-roll ($\epsilon_1,\,\epsilon_2,\,\epsilon_3\ll 1$) and ultra-slow-roll ($\epsilon_1,\,\epsilon_3\ll 1,\,\epsilon_2\simeq -6$) backgrounds one has $z''/z \simeq 2 \mathcal{H}^2$, linear perturbations in these backgrounds have identical spectra when expressed in terms of the Mukhanov-Sasaki variable $v_k = z \mathcal{R}_k$. While one does see avatars of this duality in the intermediate steps of the calculation performed above,\footnote{For example, the potential $V(X)$ of Eq.~\eqref{eq:potential:X} and the Bessel index $\nu = |3+\epsilon_2|/2$ for the mode functions in Eq.~\eqref{eq:genRsol} are identical for SR ($\epsilon_2 = 0$) and USR ($\epsilon_2 = - 6$) backgrounds.} it does not survive the non-linear mapping. 

Finally, perhaps the most significant implication of the above results is that the prospect of detecting a genuinely quantum signature of our universe's origins in cosmological observables may be less bleak than previously thought. There is, in principle, a substantial degree of ``quantumness'' encoded in the CMB. As discussed in Sec.~\ref{sec:intro}, one manifestation of this is the substantial amount of quantum entanglement between opposite Fourier modes, which nevertheless does not lead to significant quantum correlations in real space. The appearance of regions of negativity in the Wigner function instead suggests that there should exist observables sensitive to the quantum interferences present in non-slow-roll backgrounds --- though we stress that establishing the existence of negativity is only a necessary prerequisite to the identification of concrete observables able to reveal it. The latter is a separate, more arduous task which we leave to future work.

There are several natural directions along which the present work can be extended. To place them in context, it is useful to situate our approach within the landscape of $\delta N$-based methods, which are all rooted in the separate-universe picture discussed in Sec.~\ref{sec:matchingprescription}. Within this landscape, one may broadly distinguish between the original $\delta N$ formalism, which describes a classical closed system, and its stochastic generalisation, which remains classical but treats the long-wavelength sector as an open system by coarse-graining at a fixed physical scale and incorporating backreaction from short-wavelength modes as they join the system. The framework developed here can be viewed as a quantum version of the separate-universe picture underlying the $\delta N$ formalism, since we work at fixed comoving scale within a closed, but fully quantum, system. This perspective immediately suggests two extensions:
\begin{itemize}
    \item \textbf{Gradient corrections:} The separate-universe approach corresponds to the lowest-order approximation in a gradient expansion and therefore breaks down when these gradient terms become important, such as during sudden transitions away from the slow-roll attractor~\cite{Leach:2001}. Deviations from slow-roll behaviour are necessary in single-field models seeking to enhance small-scale power, a scenario of much theoretical interest and one in which genuinely quantum effects, such as interferences in the Wigner function, are expected to play a significant role. In this work, we have focused on the USR phase, but the inclusion of gradient corrections would enable the embedding of this USR phase in a realistic scenario that includes transitions from, and to, the SR attractor. Extensions of the classical and stochastic $\delta N$ formalisms incorporating gradient corrections have recently been explored in~\cite{Jackson:2023,Artigas:2024ajh,Briaud:2025ayt}. Developing an analogous extension within our fully quantum framework would be a natural next step. The generalisation to finite-$k$ is also necessary to compute the full momentum-space Wigner function, from which $n$-point functions (e.g. bispectrum, trispectrum in various configurations) could be constructed. This is also likely a prerequisite for identifying candidate observables.

    \item \textbf{Open-system effects:} Our analysis has thus far treated the system as closed and neglected environmental decoherence, which is expected to play an important role in the classicalisation of inflationary perturbations. A more realistic treatment should instead model the system as open, either through stochastic effects\footnote{This can be accomplished by coarse graining at a fixed physical -- rather than comoving -- scale.} associated with modes crossing into the system, or through coupling to an external bath representing other environmental degrees of freedom, such as entropic perturbations. One avenue towards investigating these effects without committing to a specific microphysical realisation of the environment is the open EFT of inflation~\cite{Salcedo:2024smn}, which extends the EFT framework to incorporate dissipation, noise, and decoherence. The master equations for the density matrix that these open theories provide can be unravelled into stochastic Schr\"odinger equations~\cite{Christie:2025knc}, and computing the Wigner function along their realisations should allow one to monitor environmental effects on quantum interferences.
\end{itemize}

\noindent Understanding how such effects modify the Wigner negativity observed here remains an important open question. Nevertheless, the fact that such pronounced negativity occurs already in one of the simplest well-motivated non-slow-roll backgrounds is very encouraging. Further, that this negativity occurs so close to the origin in phase space when cosmological fluctuations are large suggests that the associated quantum interference effects may be accessible to cosmological observables involving large-density objects such as ultra-compact halos or primordial black holes. Our results demonstrate that going beyond linear order and slow-roll evolution opens up a qualitatively new regime in which the quantum nature of primordial fluctuations may manifest. Moreover, the framework developed here provides an ideal theoretical setting in which these and further questions can be systematically explored.

\acknowledgments
The authors would like to thank G. Barenboim, H. Firouzjahi, D. Kaiser, and B. Nikbakht for useful conversations. AI would also like to thank M. Von Drasek for providing computational resources. AI is supported by NSF Grant PHY-2310429, Simons Investigator Award No.~824870, DOE HEP QuantISED award \#100495, the Gordon and Betty Moore Foundation Grant GBMF7946, and the U.S.~Department of Energy (DOE), Office of Science, National Quantum Information Science Research Centers, Superconducting Quantum Materials and Systems Center (SQMS) under contract No.~DEAC02-07CH11359.
Some of the computing for this project was performed on the Sherlock cluster. The authors would like to thank Stanford University and the Stanford Research Computing for providing computational resources and support that contributed to these research results.

\appendix
\section{Canonical transformations and boundary terms}\label{app:boundaryterms}

Here we demonstrate how one may remove total time derivative terms in the action by making appropriate canonical transformations. We first review the formalism developed in~\cite{goldstein2002classical,Grain:2019vnq,Braglia:2024} to determine new phase-space variables which simplify a Hamiltonian system. Then, as an example, we demonstrate how to use this formalism to eliminate the total time derivative in Eq.~(\ref{eq:S:before:canonical:transform}).

\subsection{Canonical transformations}

Consider the field $\chi(t,\mathbf{x})$ and its conjugate momentum $\pi_\chi(t,\mathbf{x})$ which satisfy the following Poisson bracket 
\begin{equation}
    \{ \chi(t,\mathbf{x}), \pi_\chi(t,\mathbf{y}) \} = \delta^{(3)} (\mathbf{x} - \mathbf{y}) \,,
\end{equation}
for all $t$. 
The Hamiltonian density $\mathsf{H}(\chi,\pi_\chi,t)$ governs the dynamics of the system by defining the equations of motion
\begin{equation}
    \dot{\chi} = \frac{\partial \mathsf{H}}{\partial \pi_\chi} \,, \,\,\, \dot{\pi}_\chi = - \frac{\partial \mathsf{H}}{\partial \chi} \,.
\end{equation}
A canonical transformation is a transformation of phase-space variables which preserves the Poisson bracket. That is, we define new canonical variables $\tilde{\chi}$, $\tilde{\pi}_\chi$ which obey the same Poisson bracket structure 
\begin{equation}
    \{ \tilde{\chi}(t,\mathbf{x}), \tilde{\pi}_\chi(t,\mathbf{y}) \} = \delta^{(3)} (\mathbf{x} - \mathbf{y}) \,.
\end{equation}
The new Hamiltonian density $\tilde{\mathsf{H}}$ determines the evolution of the new variables via 
\begin{equation}
    \dot{\tilde{\chi}} = \frac{\partial \tilde{\mathsf{H}}}{\partial \tilde{\pi}_\chi} \,, \,\,\, \dot{\tilde{\pi}}_\chi = - \frac{\partial \tilde{\mathsf{H}}}{\partial \tilde{\chi}} \,.
\end{equation}

In practice, one can find the new Hamiltonian density $\tilde{\mathsf{H}}$ with the aid of the generating function $\mathcal{F}_i$. There are four types of generating function, corresponding to the four possible types of canonical transformations. We will focus on the Type II transformation with generating function $\mathcal{F}_2 = \mathcal{F}_2(\chi, \tilde{\pi},t)$, since this will be the type required to eliminate the boundary terms in both Eq.~(\ref{eq:S:before:canonical:transform}) and Eq.~(\ref{eq:Lwithtotderiv}). In Type II transformations, the original momentum $\pi_\chi$ and new canonical variable $\tilde{\chi}$ are expressed in terms of the new momentum $\tilde{\pi}_\chi$ and original variable $\chi$ and determined by the generating function as
\begin{equation}\label{eq:genfuncrel}
    \pi_\chi = \frac{\partial \mathcal{F}_2}{\partial \chi} \,, \quad \tilde{\chi} = \frac{\partial \mathcal{F}_2}{\partial \tilde{\pi}_\chi} \,.
\end{equation}
The new Hamiltonian density is found by demanding that the least action principle derived from the action corresponding to the original Hamiltonian remain invariant under the canonical transformation. This leads to
\begin{equation}\label{eq:newH}
    \tilde{\mathsf{H}}(\tilde{\chi}, \tilde{\pi}_\chi) = \mathsf{H} \big( \chi(\tilde{\chi}, \tilde{\pi}_\chi), \pi_\chi(\tilde{\chi}, \tilde{\pi}_\chi) \big) + \frac{\partial \mathcal{F}_2}{\partial t} \bigg|_{\chi(\tilde{\chi}, \tilde{\pi}_\chi),\tilde{\pi}_\chi} \,.
\end{equation}

\subsection{Example: simplifying Eq.~(\ref{eq:S:before:canonical:transform})}

As an example, we demonstrate how to eliminate the total-derivative term in Eq.~(\ref{eq:S:before:canonical:transform}), which corresponds to the Lagrangian
\bea 
    \mathcal{L}= & \Mp^2 a^3(t)H^2(t)\epsilon_1(t+\chi)\left[ \dot{\chi}^2  - \frac{1}{a^2(t)}(\partial \chi)^2 \right]{\color{blue}-\frac{\dd}{\dd t}\bigg[2\Mp^2 a^3(t)H(t+\chi)\bigg]} \,,
\eea 
where, hereafter, the contributions from the total-derivative term are displayed in blue. The momentum conjugate to $\chi$ is defined as $\pi_\chi = \frac{\partial \mathcal{L}}{\partial \dot{\chi}}$, leading to
\bea 
    \pi_\chi = & 2\Mp^2 a^3(t)H^2(t) \epsilon_1(t+\chi) \left({\color{blue}1}+\dot{\chi}\right) ,
\eea
where we recall that in the decoupling limit we work at leading order in $\epsilon_1$. The Hamiltonian can then be obtained by the Legendre transform $\mathsf{H} = \pi_\chi \dot{\chi} - \mathcal{L}$, 
\bea
    \mathsf{H} = & \frac{\pi_\chi^2}{4\Mp^2 a^3(t)H^2(t)\epsilon_1(t+\chi)} + \Mp^2 a(t) H^2(t)\epsilon_1(t+\chi)\left(\partial\chi\right)^2
    \\ &
    {\color{blue} - \Mp^2 a^3(t)H^2(t)\epsilon_1(t+\chi)-\pi_\chi +6 \Mp^2 a^3(t) H(t) H(t+\chi) } \,.
\eea
Since the Hamiltonian is quadratic in $\pi_\chi $ but highly non-linear in $\chi$, let us look for a canonical transformation where only the momentum is redefined. In addition, since $\chi$ and $t$ only appear through the combinations $t$ and $t+\chi$ in the Hamiltonian, let us consider a canonical transformation of the form
\bea 
    \chi=\tilde{\chi} \,, \quad \pi_\chi  = \tilde{\pi}_\chi +{\color{magenta}h(t) g\left(t+\tilde{\chi}\right)} \,.
\eea 
Hereafter, we display in magenta the terms arising from the canonical transformation. From Eq.~(\ref{eq:genfuncrel}), such a transformation follows from the Type II generating function
\bea
    \mathcal{F}_2\left(\chi,\tilde{\pi}_\chi,t\right) = \chi \tilde{\pi}_\chi +{\color{magenta} h(t) G\left(t+\chi\right)} \,,
\eea
where $\partial_\chi G = g$. Making use of Eq.~(\ref{eq:newH}), the new Hamiltonian density reads
\bea\label{eq:Hamiltonian:tranformed}
    \tilde{\mathsf{H}} = &\frac{\tilde{\pi}_\chi^2}{4\Mp^2 a^3(t)H^2(t)\epsilon_1(t+\tilde{\chi})} +\Mp^2 a(t) H^2(t)\epsilon_1(t+\tilde{\chi}) \left(\partial\tilde{\chi}\right)^2
    \\ &
    +\tilde{\pi}_\chi \left[{\color{magenta} \frac{h(t)g\left(t+\tilde{\chi}\right)}{2\Mp^2 a^3(t) H^2(t)\epsilon_1\left(t+\tilde{\chi}\right)}} - {\color{blue}1}\right]
    \\ &
    +\Mp^2 a^3(t) H^2(t) \epsilon_1\left(t+\tilde{\chi}\right) 
    \left\lbrace{\color{magenta} \left[\frac{h(t)g\left(t+\tilde{\chi}\right)}{2\Mp^2 a^3(t)H^2(t)\epsilon_1\left(t+\tilde{\chi}\right)}\right]^2} - {\color{blue}1}\right\rbrace \\
    & + {\color{blue} 6 \Mp^2 a^3(t) H(t) H(t+\tilde{\chi})} +{\color{magenta} \dot{h}(t)G(t+\tilde{\chi})} \,.
\eea 
From the above expression, it is clear that the functions $h$ and $g$ that allow the terms generated by the canonical transformation to cancel out those coming from the total derivative are $h(t)=2 \Mp^2 a^3(t)H^2(t)$ and $g(t+\tilde{\chi})=\epsilon_1(t+\tilde{\chi})$, for which $G(t+\tilde{\chi})=-H(t+\tilde{\chi})/H^2(t)$. We therefore conclude that, under the redefinition
\bea 
    \chi=\tilde{\chi}, \quad \pi_\chi = \tilde{\pi}_\chi +2 \Mp^2 a^3(t) H^2(t) \epsilon_1(t+\tilde{\chi}) \,,
\eea 
the Hamiltonian reduces to the first line in Eq.~(\ref{eq:Hamiltonian:tranformed}), which is the expression used in the main text.

\section{Additional plots}

\subsection{Negativity within the perturbative domain}\label{app:perturbative}

\begin{figure}[h!]
\centering
\includegraphics[width=0.55\textwidth]{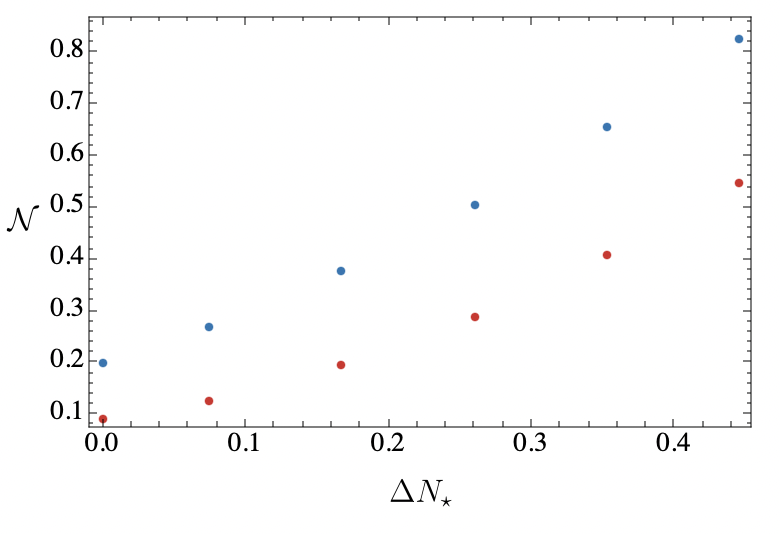}
\caption{Wigner negativity~\eqref{fig:negativity} for the reference benchmark of Eq.~(\ref{eq:RB}) (red) and a second benchmark with $(\bar{\mathcal{P}}_\mathcal{R},\sigma) = (0.28,1)$ computed by restricting the integration domain to the perturbative regime ($|\epsilon_2 H \chi_0| < 1$). Comparing this behaviour with Fig.~\ref{fig:negativity}, one sees that the trend has flipped, such that smaller scalar amplitude values $\bar{\mathcal{P}}_\mathcal{R}$ are associated with smaller negativities.
}
\label{fig:perturbative}
\end{figure}

In Fig.~\ref{fig:perturbative} we restrict the computation of Wigner negativity to the phase-space region $\vert\epsilon_2 H\chi_0\vert<1$ corresponding to curvature perturbations being indeed perturbative. Note that, since the action~\eqref{eq:action} is quadratic in $\dot{\chi}$ but highly non-linear in $\chi$, perturbativity, defined as the phase-space domain which can be properly described by the action when truncated at second order in $\chi$ and $\dot{\chi}$, only involves an upper bound on $\vert \chi\vert$. Negativity increases with $\bar{\mathcal{P}}_{\mathcal{R}}$, in agreement with the intuition that larger $\bar{\mathcal{P}}_{\mathcal{R}}$ results in stronger non-perturbative effects, and hence larger negativity.

\subsection{Varying the coarse-graining parameter}\label{app:sigma}

\begin{figure}[b!]
\centering
\includegraphics[width=0.72\textwidth]{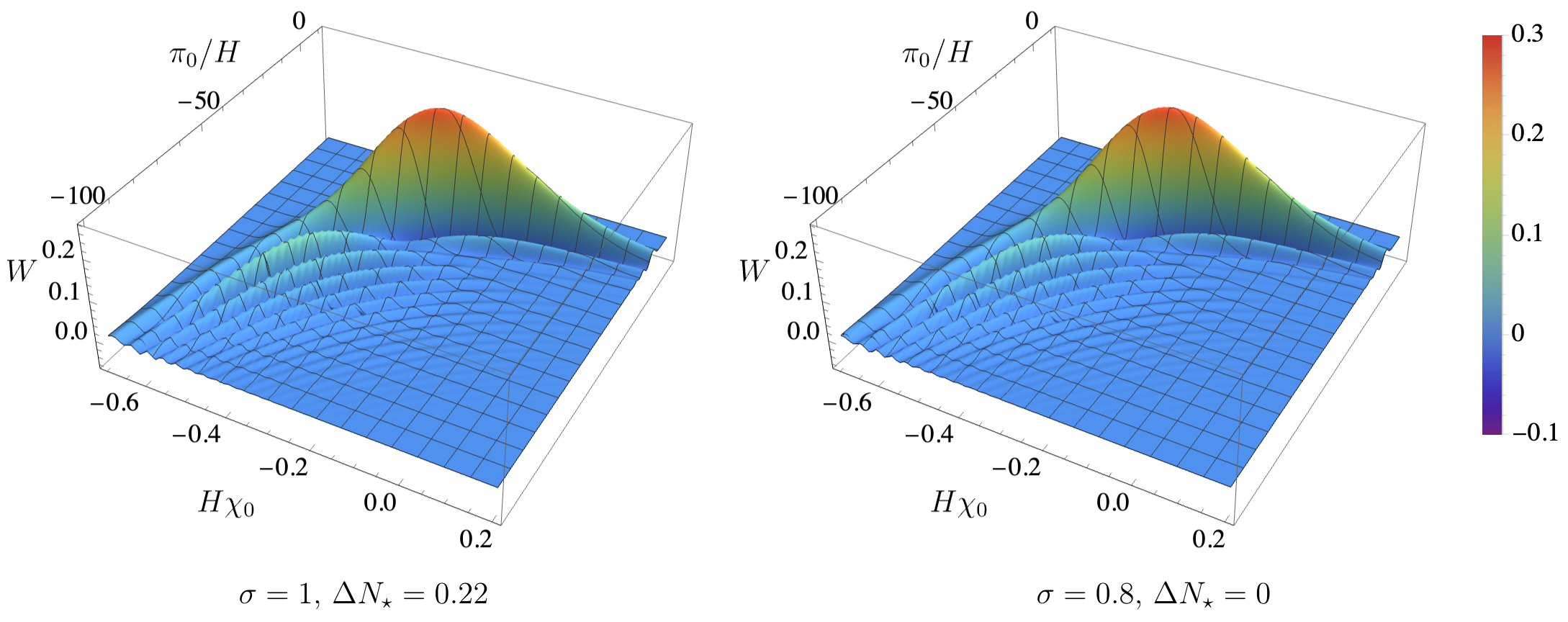}
\hspace{10cm}
\includegraphics[width=0.72\textwidth]{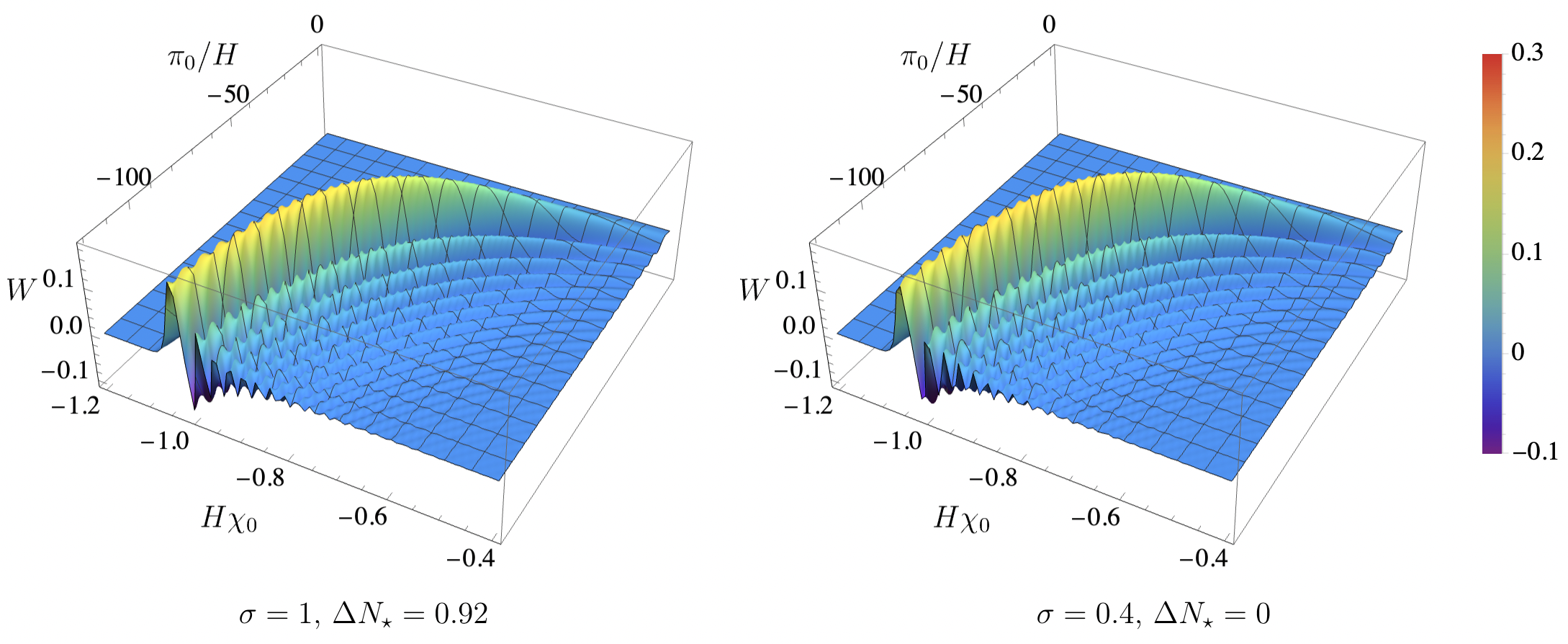}
\caption{USR Wigner function upon varying the coarse-graining parameter $\sigma$. \textbf{Top:} $W$ for $\sigma=1$ and $\Delta N_\star=-\ln(0.8)$ (left) compared with $\sigma=0.8$ and $\Delta N_\star=0$ (right). \textbf{Bottom:} $W$ for $\sigma=1$ and $\Delta N_\star=-\ln(0.4)$ (left) compared with $\sigma=0.4$ and $\Delta N_\star=0$ (right). All panels are evaluated at the same time $N$ and $\bar{\mathcal{P}}_{\mathcal{R}}$ is set to the RB value~\eqref{eq:RB}.
}
\label{fig:USRsigma}
\end{figure}

Here we check that our results are relatively robust under modest variations in the coarse-graining parameter $\sigma$. As illustrated in Fig.~\ref{fig:USRsigma} for two representative benchmarks, the differences are visually imperceptible. This behaviour is expected to persist for modestly smaller values of $\sigma$, although demonstrating this would require impractically dense sampling. The tiny residual discrepancies between the left and right panels arise because, in our separate-universe-inspired framework, non-linear effects are included only \textit{after} the matching time $N_\star = - \ln(\sigma)$. Consequently, taking $\sigma$ to be too small leads to an underestimation of their impact.

\bibliographystyle{JHEP}
\bibliography{biblio}

\end{document}